\documentclass{emulateapj}

\slugcomment{}
\usepackage{graphicx}
\usepackage{longtable}

\shorttitle{The SED Properties and SFHs of $B_W$ dropouts}
\shortauthors{Lee et al. }
\begin{document}
\def\hh{\, h^{-1}}
\newcommand{\ie}{$i.e.$,}
\newcommand{\wth}{$w(\theta)$}
\newcommand{\mpc}{Mpc}
\newcommand{\xir}{$\xi(r)$}
\newcommand{\Lya}{Ly$\alpha$}
\newcommand{\Lyb}{Lyman~$\beta$}
\newcommand{\Hb}{H$\beta$}
\newcommand{\msun}{M$_{\odot}$}
\newcommand{\hmsun}{$h^{-1}$M$_{\odot}$}
\newcommand{\sfr}{M$_{\odot}$ \text{yr}$^{-1}$}
\newcommand{\dnsty}{$h^{-3}$Mpc$^3$}
\newcommand{\za}{$z_{\rm abs}$}
\newcommand{\ze}{$z_{\rm em}$}
\newcommand{\cmtwo}{cm$^{-2}$}
\newcommand{\nhi}{$N$(H$^0$)}
\newcommand{\degpoint}{\mbox{$^\circ\mskip-7.0mu.\,$}}
\newcommand{\halpha}{\mbox{H$\alpha$}}
\newcommand{\hbeta}{\mbox{H$\beta$}}
\newcommand{\hgamma}{\mbox{H$\gamma$}}
\newcommand{\kms}{\,km~s$^{-1}$}      
\newcommand{\minpoint}{\mbox{$'\mskip-4.7mu.\mskip0.8mu$}}
\newcommand{\mv}{\mbox{$m_{_V}$}}
\newcommand{\Mv}{\mbox{$M_{_V}$}}
\newcommand{\peryr}{\mbox{$\>\rm yr^{-1}$}}
\newcommand{\secpoint}{\mbox{$''\mskip-7.6mu.\,$}}
\newcommand{\sqdeg}{\mbox{${\rm deg}^2$}}
\newcommand{\squig}{\sim\!\!}
\newcommand{\subsun}{\mbox{$_{\twelvesy\odot}$}}
\newcommand{\et}{{\it et al.}~}
\newcommand{\er}[2]{$_{-#1}^{+#2}$}
\def\h50{\, h_{50}^{-1}}
\def\hbl{km~s$^{-1}$~Mpc$^{-1}$}
\def\ltsima{$\; \buildrel < \over \sim \;$}
\def\simlt{\lower.5ex\hbox{\ltsima}}
\def\gtsima{$\; \buildrel > \over \sim \;$}
\def\simgt{\lower.5ex\hbox{\gtsima}} 
\def\arcs{$''~$}
\def\arcm{$'~$}
\newcommand{\wu}{$U$}
\newcommand{\wb}{$B_{435}$}
\newcommand{\wv}{$V_{606}$}
\newcommand{\wi}{$i_{775}$}
\newcommand{\wz}{$z_{850}$}
\newcommand{\hmpc}{$h^{-1}$Mpc}
\newcommand{\lm}{$L$--$M$}
\newcommand{\ws}{$\mathcal{S}$}
\newcommand{\wm}{$\mathcal{M}$}
\newcommand{\sm}{$\mathcal{S}$-$\mathcal{M}$}
\newcommand{\medianLM}{$\tilde{\mathcal{L}}(M)$}
\newcommand{\mf}{$\phi_\mathcal{M}$}

\title{The Average Physical Properties and Star Formation Histories \\of the UV-Brightest Star-forming Galaxies at $z\sim3.7$}
\author{Kyoung-Soo Lee\altaffilmark{1,2}, Arjun Dey\altaffilmark{3}, Naveen Reddy\altaffilmark{3,4}, Michael J.~I.~Brown\altaffilmark{5}, Anthony H.~Gonzalez\altaffilmark{6}, Buell  T. Jannuzi\altaffilmark{3}, Michael C.~Cooper\altaffilmark{7,8}, Xiaohui Fan\altaffilmark{7}, Fuyan Bian\altaffilmark{7},  Eilat Glikman\altaffilmark{1,9},  Daniel Stern\altaffilmark{10}, Mark Brodwin\altaffilmark{11,12}, Asantha Cooray\altaffilmark{13}}
\altaffiltext{1}{Yale Center for Astronomy and Astrophysics, Department of Physics, Yale University,  New Haven, CT 06520}
\altaffiltext{2}{Jaylee and Gilbert Mead Fellow}
\altaffiltext{3}{National Optical Astronomy Observatory, Tucson, AZ 85726}
\altaffiltext{4}{Hubble Fellow}
\altaffiltext{5}{Monash University, Clayton, Victoria 3800, Australia}
\altaffiltext{6}{Department of Astronomy, University of Florida, Gainesville, FL 32611}
\altaffiltext{7}{Steward Observatory, University of Arizona, Tucson, AZ 85721}
\altaffiltext{8}{Spitzer Fellow}
\altaffiltext{9}{National Science Foundation Fellow}
\altaffiltext{10}{Jet Propulsion Laboratory, California Institute of Technology, MS 169-327, 4800 Oak Grove Drive, Pasadena, CA 91109}
\altaffiltext{11}{Harvard-Smithsonian Center for Astrophysics,  Cambridge, MA 02138}
\altaffiltext{12}{W. M. Keck Postdoctoral Fellow}
\altaffiltext{13}{University of California, Irvine, CA 92697}

\begin{abstract}
We investigate the average physical properties and star formation histories (SFHs) of the most UV-luminous star-forming galaxies at $z\sim3.7$. Our results are based on the average spectral energy distributions (SEDs), constructed from stacked optical to infrared photometry, of a sample of the 1,913 most UV-luminous star-forming galaxies found in 5.3 square degrees of the NOAO Deep Wide-Field Survey. We find that the shape of the average SED in the rest-optical and infrared is fairly constant with UV luminosity: i.e., more UV luminous galaxies are, on average, also more luminous at longer wavelengths.  In the rest-UV,  however, the spectral slope $\beta$ ($\equiv~d{\rm log}F_\lambda/d{\rm log}\lambda$; measured at $0.13\mu\rm{m}<\lambda_{\rm{rest}}<0.28\mu\rm{m}$) rises steeply with the median UV luminosity from $-1.8$ at $L\approx L^*$ to $-1.2$  ($L\approx 4-5L^*$). We use population synthesis analyses to derive their average physical properties and find that: (1) $L_{\rm{UV}}$, and thus star-formation rates (SFRs), scale closely with stellar mass such that more UV-luminous galaxies are  more massive; (2) The median ages indicate that  the stellar populations are relatively young (200-400~Myr) and show little correlation with UV luminosity;  and (3) More UV-luminous galaxies are dustier than their less-luminous counterparts, such that $L\approx 4-5L^*$ galaxies are extincted up to $A(1600)=2$ mag while $L\approx L^*$ galaxies have $A(1600)=0.7-1.5$ mag. We argue that the average SFHs of UV-luminous galaxies are better described by models in which SFR increases with time in order to simultaneously reproduce the tight correlation between the UV-derived SFR and stellar mass, and their universally young ages. We demonstrate the potential of measurements of the SFR-$M_*$ relation at multiple redshifts to discriminate between simple models of SFHs. Finally, we  discuss the fate of these UV-brightest galaxies in the next $1-2$~Gyr and their possible connection to the most massive galaxies at $z\sim2$.
\end{abstract}
  

\section{Introduction}
Constraining the main mode of star formation at high redshift is of fundamental importance to the theory of galaxy formation. 
Significant progress has been made in characterizing some of the general physical properties of the star-forming galaxy population at high redshift. Measurements have been made of the clustering, spectral energy distributions (SEDs), and kinematics of these galaxies \citep[e.g.,][]{shapley01, shapley03, papovich01, pentericci07, lee06, lee09, stark09, vanzella09}. Despite these efforts, major questions remain about the details of how these galaxies form their stars. We do not yet know whether their star formation history is dominated by a series of short but intense ``bursts'' or a long continuous accretion of gas converted into stars.
Constraints on the star formation history of galaxy populations can bear on the availability of fuel for star formation, and hence their gas accretion history. For example, bursts may point to merger-induced processes which deliver galaxy-sized parcels of gas \citep[e.g.][]{mihos96, kolatt99, somerville01},
whereas continuous star formation might suggest a more stable fuel delivery system, such as ``cold-mode accretion" \citep[e.g.,][]{keres08, dekel09}. 
Observations at $z=0-1$ \citep[e.g.,][]{noeske07, elbaz07} and a limited set of observations at redshift out to $z\sim2$ \citep{genzel06, daddi07a}, however, show some inconsistencies with theoretical predictions at high redshift \citep{dave08}. For example, the studies of faint star-forming galaxies at $z>3$ seem to favor short duty cycles unlike theoretical expectations \citep [e.g.,][]{lee09, stark09}. 

It is currently unclear whether such discord requires changes to the theory (e.g., implementing luminosity-dependent galaxy formation) or is the result of observational biases inherent in our incomplete census of galaxies at high redshift. This is because most recent efforts have been focused on samples derived from the small-area, deep surveys \citep[e.g., GOODS, HUDF:][]{sawicki06, bouwens07}, which while excellent at 
identifying sub-$L^*$ galaxies, lack the cosmic volume to contain significant samples of the most luminous and massive galaxies. While the UV-faint galaxies likely dominate the cosmic star formation rate density \citep[e.g.,][]{madau96, reddy09}, their measured star-formation rates, space densities and clustering amplitudes \citep[e.g.,][]{lee06, bouwens07, reddy09} suggest that they are not likely to be the progenitors of the most massive galaxies seen at later epochs \citep{quadri07, elsner08, marchesini09}. In contrast, the UV-brightest galaxies at $z>3$ may provide a more plausible progenitor population for the most massive galaxies ($M_{\rm{star}}\gtrsim \rm{a~few} \times 10^{11}M_\odot$), especially if they are able to sustain their observed star-formation rates for an extended duration ($\gtrsim$ 1 Gyr).  

In this paper, we define a  large sample ($1913$) of the most UV luminous galaxies at $z>3$, and investigate their average properties with the goal of determining their stellar masses, star formation histories, and evolutionary descendants. 

We use $(\Omega, \Omega_\Lambda, \sigma_8, h_{100})=(0.28, 0.72, 0.9, 0.72)$. Magnitudes are given in AB system \citep{oke83} unless noted otherwise.
 
\section{Data and Galaxy Sample}

Our sample of 1,913 super-$L^*$ star-forming galaxies at $z\approx 3.7$ is selected from the 9.3~deg$^2$ Bo\"otes field of the NOAO Deep Wide-Field Survey \citep[NDWFS, hereafter:][]{jannuzi_dey99}. The NDWFS dataset used in this work includes the optical $B_WRI$ imaging \citep[][available from the NOAO Science Archive]{jannuzi_dey99} taken with the MOSAIC camera, near-infrared $JHK_S$ taken with the NEWFIRM camera on the Mayall 4m telescope (A.~Gonzalez et al., in prep), and the {\it Spitzer} IRAC data from the Spitzer Deep Wide-Field Survey \citep[3.6, 4.5, 5.8, 8.0 $\mu$m;][available from the Spitzer Heritage Archive]{ashby09}. Where available, we also use $U$-band (``$U$-spec'' filter) data  taken with the Large Binocular Telescope (F.~Bian et al., in prep) to check whether galaxies in our sample are detected. Because the optical data were taken over multiple pointings, the depth and seeing vary significantly over the entire Bo\"otes field. In order to ensure robust photometric measurements and candidate selection, we restrict our study to areas where the seeing in the $B_W$, $R$ and $I$ bands has a Moffat profile with a full width at half maximum MFWHM$\le 1.3\arcsec$, with a best-fit Moffat parameter $\beta<2.5$. Using these criteria, we exclude five subfields of the NDWFS data (NDWFSJ1426p3421, NDWFSJ1428p3456, NDWFSJ1428p3531, NDWFSJ1431p3346, and NDWFSJ1434p3456). The seeing and photometric depths of the optical data in all subfields are found on the survey website\footnote{\tt http://www.noao.edu/noao/noaodeep/}. After excluding these subfields and the masked area, the effective area of our sample is $\approx$5.3 degree$^2$. 


Our sample is selected by applying a Lyman-break color-selection technique to the $B_WRI$ data. 
The Lyman-break technique is designed to identify a UV bright star-forming galaxy with a strong spectral break (at wavelength $\lambda\le 1216$ \AA) resulting from absorption by the intervening Ly$\alpha$ forest. At redshifts $3.3<z<4.3$, this break lies between the $B_W$ and $R$ bands. Following numerous studies in the literature \citep[e.g.,][]{steidel99, mauro04b, bouwens07}, we simulated the expected colors of typical star-forming galaxy SEDs at a range of redshifts and defined the following color criteria:
\begin{eqnarray}
(B_W-R) \ge 1.2+3.0 \times (R-I) ~~~~~ (B_W-R) \ge 2.0~~ \nonumber\\
 (R-I) \ge -0.3 ~~~~~ S/N(R) \ge 3 ~~~~~ S/N(I) \ge 7~~~~~~~
\end{eqnarray}
In order to only include objects with robust photometric measurements, we excluded candidates lying within masked regions around bright stars (including  diffraction spikes around them), large galaxies, and the edge of each subfield image.   We also excluded objects that are heavily blended with adjacent sources and stars based on the SExtractor CLASS\_STAR indices \citep{bertina96}.  Of 108 objects that satisfy the criterion ${\rm CLASS\_STAR} >  0.95$, 78 objects (72\%) are brighter than $I=23.0$ where  the stellarity index can be measured reliably.  Furthermore, we use optical-infrared colors \citep[$R-I$ vs. $I-3.6 \mu$m define a stellar locus well separated from galaxies; see Figure 8b in ][]{brown07} to further exclude stars whenever possible. We also visually examined the LBT $U$-band data and exclude any of candidates which are detected; the $U$-band samples light below the Lyman continuum break ($\lambda=912$ \AA) at redshifts $z\gtrsim3.4$, and hence any candidate detected in $U$ is unlikely to be at the redshift of interest.  
Very few ($<$1\%) of the sources are identified in the U-band (with detections defined as 3$\sigma$ or higher, corresponding to $U\le 26$ mag). After excluding these individual detections, we stacked the resulting sample and confirmed that the median candidate is also not detected in the (stacked) $U$-band image (see Figure \ref{stacked_image}).
We note that less than 10\% of our candidates are formally detected in the IRAC channel 1 and 2 (3.6$\mu$m), and less than 1\% in channel 3 and 4. While we do not exclude any object based on the IRAC photometry, this is reassuring because the relatively shallow depth of the {\it Spitzer} data of the Bo\"otes field is unlikely to warrant detection of most high-redshift sources except for very massive ones (see later). The 5-sigma depths of the SDWFS data are 22.74, 22.24, 20.39, 20.25 mag (AB) for IRAC channel 1, 2, 3, and 4, respectively \citep{ashby09}.

\begin{figure*}[t]
\epsscale{1.1}
\plottwo{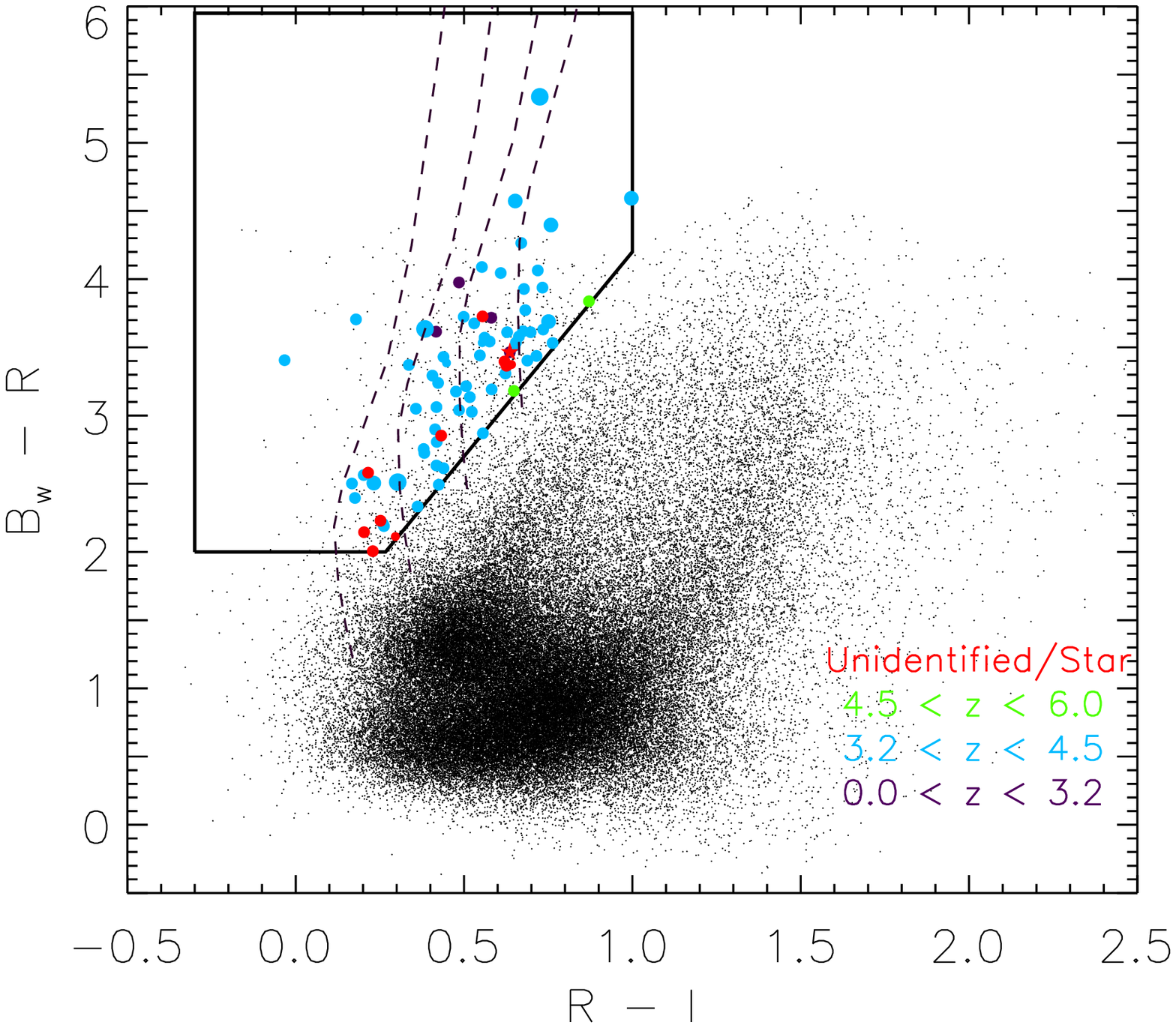}{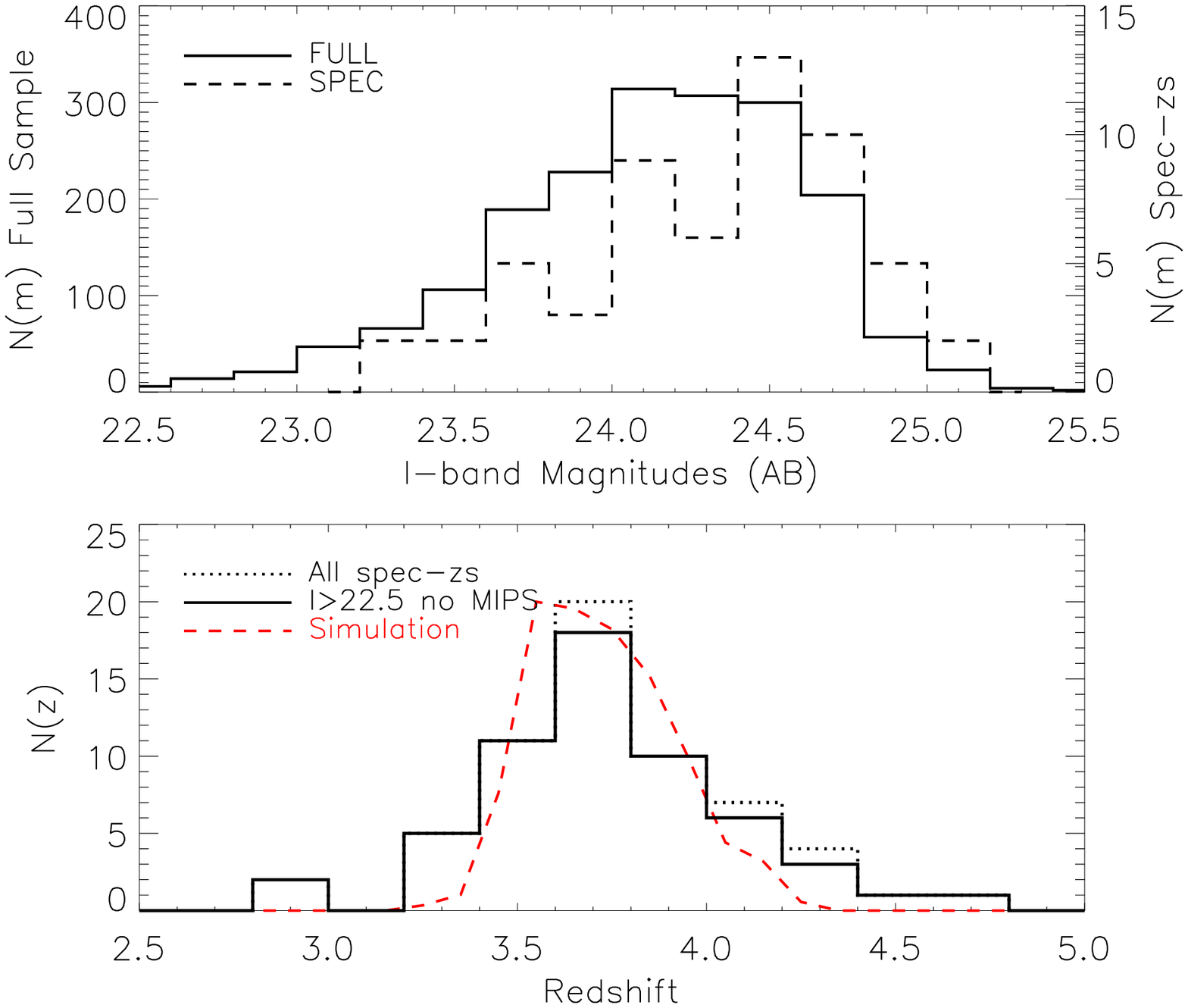}
\caption[redshift]{{\it Left:} The $B_WRI$ color-color plane with the adopted color selection criteria for $z\ge 3.3$ candidates denoted by the black lines. The black dots represent a subset of all NDWFS galaxies. Colored symbols indicate the sources with spectroscopic redshifts, demonstrating the relatively clean selection of high-redshift targets. Dashed lines show the theoretical tracks of star-forming galaxies at $z\ge3.3$ ($z=3.3$ at the bottom) for four reddening parameters ($E(B-V)$=0.0, 0.15, 0.30, 0.45).  {\it Right:} The magnitude distribution of the full sample (``FULL'': solid) and spectroscopic sample (``SPEC'': dashed) are shown (top). 
The bottom panel shows the redshift distributions of the entire spectroscopic sample (dotted line) and the subset containing only the fainter galaxies with $I\ge 22.5$ (solid line).
We adopt the latter as the representative redshift distribution for the full photometric sample. The expected redshift distribution of the full sample from our simulations is in good agreement with the observed one (dashed line). 
}
\label{plot_redshift}
\end{figure*}

The final number of candidates is 1,913, corresponding to a surface density of $\approx 0.10$ arcmin$^{-2}$ over the 5.3 degree$^2$ area. As shown in Figure \ref{plot_redshift} ({\it top right}), our sample contains a significant number of very bright sources  -- for example,  236 sources brighter than $I_{AB}=23.5$; these provide ideal targets for detailed studies of galaxy properties such as their stellar populations, interstellar media, and kinematics. 

\section{Spectroscopic Sample}
\begin{figure*}[t]
\epsscale{1.0}
\plotone{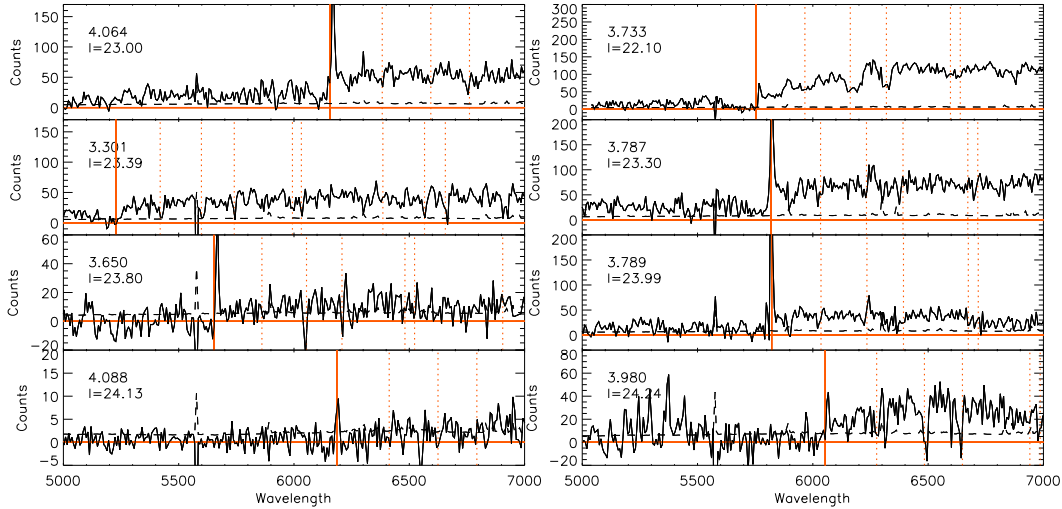}
\caption[redshift]{Example DEIMOS spectra of $B_W$-band dropout candidates. The spectra are ordered in increasing $I$-band magnitudes from top to bottom. Solid vertical lines mark the location of Ly$\alpha$  while dotted vertical lines mark, from left to right, Si II 1260 \AA, OI+Si II 1302 \AA, C II 1335 \AA,  Si IV 1394,1403 \AA, Si II 1527 \AA, C IV 1548,1551 \AA\ in absorption, respectively. }
\label{plot_spectra}
\end{figure*}

In order to characterize the redshift distribution and contamination rates by low-redshift interlopers and Galactic stars, redshifts for a small sample of candidates were obtained using the Deep Imaging Multi-Object Spectrograph spectrograph \citep[DEIMOS; ][]{deimos_ref} on the Keck II Telescope of the W.~M.~Keck Observatory on U.T. 2010 May 15. Three masks, containing a total of 66 candidates, were observed with $2.0-2.5$ hours each using the standard DEIMOS setup with the 600 l/mm grating ($\lambda_{\rm blaze} = 7500$\AA) centered on $\sim 7600$  \AA. Mask positions were chosen to maximize the number of primary targets (secure $B_W$-band dropout candidates; ${\rm S/N} \geq 8$) and secondary targets (fainter candidates; $5 \leq {\rm S/N} < 8$) within the DEIMOS field of view. The spectra were reduced using the DEIMOS pipeline developed by the DEEP2 survey team at UC Berkeley (M.~Cooper et al., in prep). While the details of all our spectroscopic observations will be presented elsewhere, we show typical 1D spectra of our candidates in Figure \ref{plot_spectra}.  In addition to these observations, we compiled existing spectroscopic redshifts from the past Keck observations by Spinrad, Soifer and collaborators as well as those obtained from the AGN and Galaxy Evolution Survey (AGES: C.~Kochanek et al., in prep).


In total, 78 sources  from our sample have been observed spectroscopically. Of these, redshifts could not be determined for 12 sources due to optical faintness or a lack of useful lines. Two of the spectroscopically confirmed sources are {\it Spitzer} MIPS 24$\mu$m sources at $z$=4.230 and 3.675 \citep{desai08}. The former is a narrow-line AGN exhibiting CIV and HeII, and the latter is a broad-line QSO. Additionally, three sources have been identified as QSOs from our DEIMOS spectra ($z$=3.835, 4.230, 4.050), and one source is a Galactic star. The total number of confirmed interlopers (i.e., those other than star-forming galaxies at $z\sim3.7$) is 5 AGN and 1 star. Hence, the rate of contamination ranges from 8\% (if all 12 unidentified sources are at high redshift: 6/78) to 23\% (if all 12 sources are at low redshift: (6+12)/78). In Figure \ref{plot_redshift} ({\it bottom right}), we show the two redshift distributions of our spectroscopic samples, one for the full sample including the MIPS sources, and the other showing only $I\ge 22.5$ galaxies without MIPS detections. 
We further note that the MIPS 24 $\mu$m sources are fairly bright ($I\lesssim 22.5$) compared to the majority of our candidates. 
Hence, we adopt the latter as the representative redshift distribution $N(z)$, as the majority of our sample is fainter than $I=22.5$. Thus, the redshift distribution is well described by the mean $\langle z\rangle=3.7$ and the standard deviation of $0.4$. 

To further validate the adopted redshift distribution, we carried out photometric simulations to estimate the expected redshift distribution of our dropout sample based on the broad-band color selection. The simulation consists of 1) creating artificial star-forming galaxies over a range of redshifts, UV luminosity, spectral slope, morphology, and size, 2) inserting them in the real images, 3)  color-selecting in an identical  manner in which the real galaxies were selected. We refer interested readers to \citet{ferguson04} and \citet{lee06} for more details. The expected redshift distribution is shown in Figure \ref{plot_redshift} ({\it bottom right}; dashed line), in good agreement with the observed one.

\section{Measurements of the Average Spectral Energy Distributions}
We investigate the average SED properties of the UV-selected star-forming galaxies by stacking our imaging in the $B_WRIJHK_S$[3.6][4.5][5.8][8.0] passbands on the positions of the 1,913 galaxies in our sample.  For a galaxy at $z=3.7$ --  the median redshift of the sample -- these bands  probe the rest-frame wavelength range $ 0.1-2 \mu$m. In order to investigate whether their UV brightness (a proxy for star formation rate) is correlated with other galaxy properties such as stellar mass or extinction, we define six subsamples ordered by their $I$-band magnitudes. In Table 1, we summarize the details of each subsample. Given the relatively narrow redshift distribution ($z=3.7\pm0.4$), the $I$-band flux provides a fairly accurate representation of the observed UV luminosity measured at the rest-frame 1700 \AA. At a fixed $I$-band flux, the UV luminosity of a source at $z=3.7$ will be higher (lower) than that at $z=3.3$ (4.1) by $\approx 0.08$ dex (assuming that the UV spectrum has the form $f_\lambda \propto \lambda^{-1.45}$).  

In order to create average SEDs of galaxies within each luminosity bin, we first constructed stacked two dimensional images for each band. To create clean stacked images in each band, unbiased by outliers and sky subtraction uncertainties, we adopted the following procedure. (1) We first convolved all the images in a given band to a common PSF. (2) Then, from these convolved images, we extracted 1\arcmin$\times$1\arcmin\ image cutouts centered on each candidate. (3) Next, in order to obtain an unbiased estimate of the sky brightness level, we generated an object mask to mask all  sources present in each cutout, including the central one (IRAF {\tt objmasks} task was used to detect objects above a  threshold then several pixels were added to grow the mask ensuring  all the "light from sources and their halos are effectively excluded).  We then estimated the sky level using the modal value of the  unmasked pixels in each frame, and subtracted this value from the image (i.e., to create an image with zero sky). (4) Since the image cutouts come from survey data with different zero points, we then rescaled each cutout to a common zero point magnitude. (5) Finally, we constructed a median stacked image for each band from all the scaled, sky-subtracted cutouts of candidates within a given luminosity bin.  In constructing the median, we used object masks to exclude pixels from all objects other than the central one.  Using the mean instead of the median does not change our results significantly. Before the final step of stacking, we visually checked all the cutout images. Whenever the central source is significantly blended with foreground object, we exclude the source from stacking in order to avoid contamination to the stacking signal by interlopers. The effective number of sources used in stacking is reduced by $5-10$\% depending on passbands due to confusion. 


This procedure is repeated for each passband ($UB_WRIJHK_S$ and IRAC bands). Because the blending and crowding vary greatly with passbands,  slight adjustments  were made to the IRAF {\tt objmasks} parameters to ensure that all light from the surrounding sources are well masked without masking the majority of the  pixels in each cutout. The latter would result in substantial compromise in the effective depth of the stacked image as most pixels are masked and thus not used in the stacking. In Figure \ref{stacked_image} we show the stacked images of all six subsamples in the passbands where flux was detected. The stacked LBT $U$-band data yielded only an upper limit as no detection was made. The upper limit ranges over AB=$29.8$ - $30.9$ for all of our subsamples (see Table 1).

We measured aperture photometry from each stacked image and thus derived the average SED of each subsample. The photometry was measured in a circular aperture 5\arcsec\ diameter,, then aperture-corrected to derive total magnitude. For the optical  passbands, which are convolved to a common  Moffat profile PSF \citep[MFWHM=1.35\arcsec, $\beta$=2.5: see][]{brown07}, and for the near-infrared data, also convolved to a common PSF (MFWHM=1.35\arcsec\ for $H$ and $K_S$;  MFWHM=1.6\arcsec\ for $J$, $\beta$=2.5; A.~Gonzalez et al., in prep.), the aperture correction factors were estimated analytically. For the {\it Spitzer} IRAC data, we use the values published in \citet{ashby09}. The aperture correction factors are given in Table 1 in units of magnitudes. 

To estimate the impact of the intrinsic spread of the range of galaxy luminosities/magnitudes on the measured median value for the sample,
we carried out Monte Carlo simulations. Each time, a random 30\% of the sample was chosen and the stacking and photometric measurements were carried out as outlined above. We repeated this procedure 500 times to construct the flux distribution of a given subsample. The intrinsic scatter is estimated from this distribution assuming that the random errors follow a normal distribution (i.e., they can be scaled from the photometric error of the full stack), and is then added in quadrature to the photometric measurement error to estimate the total scatter in the stacked photometry (Table 1).   

\begin{figure}[t]
\epsscale{1.0}
\plotone{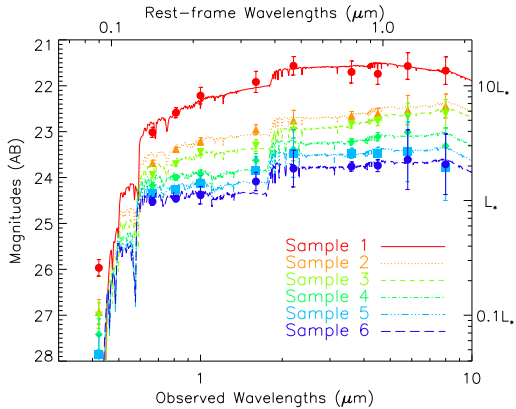}
\caption[stacked_sed]
{The average spectral energy distributions of the six UV subsamples are shown together with the best-fit population synthesis model assuming a constant star formation history. The SEDs were obtained by convolving the best-fit model with the redshift distribution $N(z)$ to account for redshift uncertainties. The $B_WRIJHK_S$ and IRAC photometry probe out to the rest-frame 2 $\mu$m at $3.3<z<4.3$. On right y-axis, we also show the characteristic UV luminosity $L^*$ at 1700 \AA\ $z\sim3$ \citep{steidel99}, corresponding to the $I$-band magnitudes (at $\lambda_{\rm{obs}}\approx 8100$ \AA, the third datapoint from left for each SED). The overall shape of the SEDs is remarkably similar to one another suggesting the regularity of their formation processes (see text). }
\label{stacked_sed}
\end{figure}

\begin{figure*}[t]
\epsscale{1.0}
\plotone{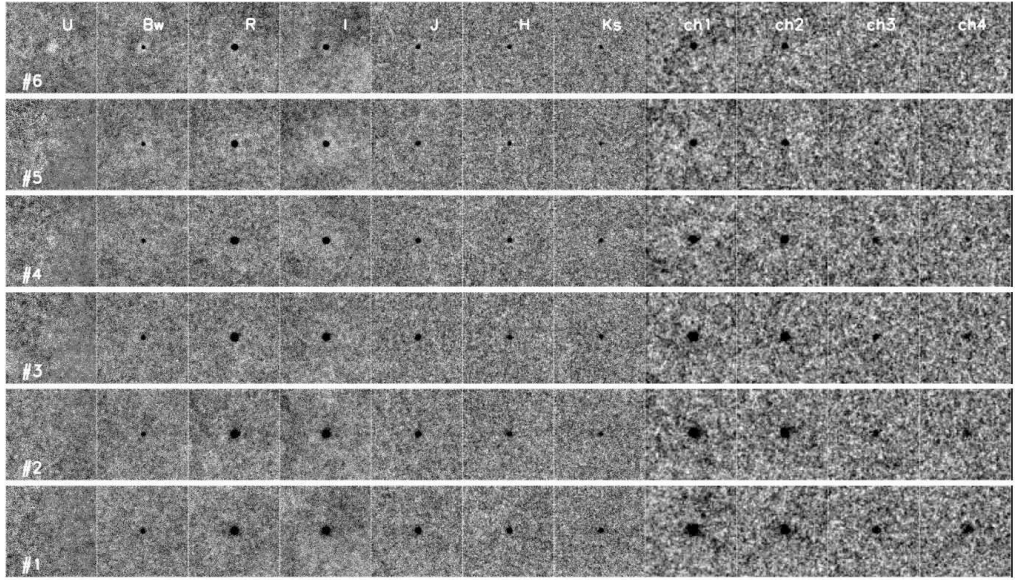}
\caption[stacked_image]
{The stacked images in the $UB_WRIJHK_S[3.6][4.5][5.8][8.0]$ passbands are shown for all six subsamples (the brightest bin on bottom). Each image has a dimension of 1\arcmin $\times$ 1\arcmin\ centered around the photometric sample. The central source is clearly detected in all passbands out to IRAC 8.0 $\mu$m with the exception of the $U$-band.}
\label{stacked_image}
\end{figure*}

As shown in Figure \ref{stacked_sed}, the average rest-frame optical SEDs of galaxies in different UV luminosity bins are strikingly similar to one another in their overall shapes. All of them share the flatness in the rest-frame optical portion of the spectrum ($f_\nu\propto\lambda^{0}$), and exhibit a weak Balmer break, which is located between the $H$ and $K_S$ bands at redshift $z=3.1-4.1$. Furthermore, it is evident that the UV-brighter galaxies are also brighter in the rest optical by roughly the same scale factor as in the UV. Despite the overall similarities, however, there are some measurable differences. In the rest-frame UV portion of the SED, the UV slope $\beta$  ($\equiv d{\rm log}F_\lambda/d{\rm log}\lambda$; measured in the observed $RIJ$ bands) is steeper with increasing UV luminosity. 
In the next section, we discuss the physical implications of our results using stellar population model fitting of the averaged SEDs.

\begin{deluxetable*}{lcccccccc}
\tabletypesize{\footnotesize}
\scriptsize
\tablecaption{Average Photometry of Star-Forming Galaxies at $z\sim3.7$}
\tablewidth{6.8in}
\tablehead{\colhead{} & \multicolumn{7}{c}{Average Photometry}  \\
\colhead{} & \multicolumn{7}{c}{-------------------------------------------------}\\
\colhead{} & $\Delta m$\tablenotemark{a} & \colhead{\#1} & \colhead{\#2} & \colhead{\#3} & \colhead{\#4} &\colhead{\#5} & \colhead{\#6} 
}
\startdata
 & -- & $19.46\le I\tablenotemark{b} < 23.06$ & $23.06\le I < 23.66$ & $23.66\le I < 23.96$ & $23.96\le I < 24.26$  & $24.26\le I < 24.46$ & $24.46\le I < 24.96$ \\
 $M_{1700}\tablenotemark{c} $ & -- & -23.30 & -22.51 & -22.18 & -21.89 & -21.63 & -21.43 \\
$N$\tablenotemark{d}             & -- & 74 & 254 & 322 & 464& 307& 495 \\
$U$ &  & $>29.84$ & $>30.51$ & $>30.63$ & $>30.83$ & $>30.61$ & $>30.87$ \\
$B_W$&  0.09 & $25.97   \pm 0.18$ &   $26.93  \pm  0.27$ &  $27.01  \pm  0.28$ &   $27.42 \pm   0.34$ &   $27.86 \pm   0.42$ &   $28.73 \pm   0.63$\\
$R$      &  0.09 & $23.01   \pm  0.11$ &   $23.68 \pm   0.06$ &  $ 23.94  \pm  0.06$ &   $24.17  \pm  0.07$ &   $24.35 \pm   0.08$ &   $24.52 \pm   0.08$\\
$I$\tablenotemark{e}        &  0.09 & $22.59  \pm  0.11$ &   $23.38   \pm 0.04$ &   $23.71 \pm   0.05$ &   $24.00 \pm   0.06$ &   $24.26  \pm  0.07$ &   $24.46  \pm 0.07$\\
$J$      &  0.14 & $22.21  \pm  0.18$ &   $23.21 \pm   0.08$ &   $23.45  \pm  0.08$ &   $23.93  \pm  0.09$ &   $24.12  \pm  0.17$ &   $24.38  \pm  0.19$\\
$H$     &  0.09 & $21.92  \pm  0.23$ &   $22.96  \pm  0.11$ &   $23.31 \pm   0.13$ &   $23.63  \pm  0.15$ &   $23.84  \pm  0.29$ &   $24.09 \pm   0.20$\\
$Ks$   &  0.09 & $21.56   \pm 0.20$ &   $22.75 \pm   0.22$ &   $22.97 \pm   0.18$ &   $23.38 \pm  0.25$ &   $23.47 \pm  0.52$ &   $23.80  \pm  0.40$\\
$3.6 \mu$m &  0.24 & $21.70 \pm  0.24$ &   $22.65 \pm   0.09$ &   $22.84 \pm   0.08$ &   $23.22  \pm  0.07$ &   $23.47 \pm   0.11$ &   $23.76  \pm  0.10$\\
$4.5 \mu$m &  0.25& $21.74 \pm   0.23$ &   $22.60  \pm  0.11$ &   $22.78  \pm  0.09$ &  $ 23.11  \pm  0.09$ &   $23.48  \pm  0.14$ &   $23.74  \pm  0.13$\\
$5.8 \mu$m &  0.38 & $21.57 \pm  0.29 $ &  $22.53 \pm   0.32$ &  $22.68  \pm  0.30$ &   $23.01  \pm  0.36$ &   $23.44 \pm   0.65$ &   $23.61 \pm   0.65$\\
$8.0 \mu$m &  0.51 & $21.67 \pm  0.30$ &  $22.44 \pm   0.26$ &  $22.61  \pm  0.31$ &  $23.31 \pm   0.39$ &   $23.78 \pm   0.72$ &   $23.71  \pm  0.66$\\
\enddata

\tablenotetext{a}{Aperture correction factor applied to the photometry measured in a 5\arcsec\ diameter to obtain the total magnitude listed in the Table. }
\tablenotetext{b}{The Kron-like total magnitude in the $I$-band \citep[SExtractor MAG\_AUTO:][]{kron80, bertina96}}
\tablenotetext{c}{Absolute magnitude at 1700 \AA\ assuming redshift $z=3.7$}
\tablenotetext{d}{The number of the photometric candidates used for stacking }
\tablenotetext{e}{$I$-band photometry measured in a 5\arcsec\ diameter with the aperture correction (see \S 3)}
\end{deluxetable*}

\section{Average Physical Properties}

\subsection{Stellar Population Synthesis Analyses}

We determine the average physical properties, specifically the stellar mass, age, and reddening, of our samples by fitting the \citet{bc03} stellar population synthesis models to the stacked photometry. We assume the Chabrier (2003) initial mass function (IMF), the \citet{calzetti00} form for the effective dust extinction law, and consider only solar metallicity populations. We model the star formation histories (SFH) to be constant, exponentially declining (SFR $\propto e^{-t/\tau}$) with a wide range of $\tau$ values (10 Myr - 5 Gyr), or linearly increasing. The stellar age is allowed to vary between 50 Myr and the age of the Universe at the redshift $z$ where the fitting is performed (i.e., $\approx1.66$ Gyr at $z\approx3.7$). The minimum age of 50 Myr is set based on dynamical timescale arguments when considering the measurements of half-light radii and  velocity dispersions of galaxies at high redshift \citep{bouwens04c, erbetal06b}.  While these assumptions for age ranges, IMF, metallicity, and extinction may not be accurate descriptions of the properties of high-redshift galaxies (and may even vary with galaxy mass and/or luminosity), our main interest lies in comparing the relative average properties of galaxies in different UV luminosity bins. Our current study is limited by the nature of our data, which is confined to broad band photometry; more details will emerge only from deep spectroscopic observations, which is now within grasp for the bright candidates within our sample. 

We estimate the uncertainties on the derived physical properties that result from the finite width of the redshift distribution using a Monte Carlo approach. We randomly draw a redshift from the observed $N(z)$ distribution ({\it right bottom} of Figure \ref{plot_redshift}), assign it to the observed average SED, find the best-fit parameters at each time, and repeat this procedure 100 times. 
Although we used the observed spectroscopic redshift distribution for these analyses, we also simulated the expected redshift distribution given our adopted UV color selection criteria and a large range of theoretical galaxy SEDs. The two distributions agree with each other reasonably well (Figure \ref{plot_redshift}), and we have verified that using one or the other does not affect our main results. 

\subsection{Luminosity, Mass and Age}

Given the narrow redshift distribution, the galaxies in our sample range in rest-frame UV luminosity from $M_{1700}\approx -21.43$ to $-23.30$ (see Table 1). 
Our sample therefore consists primarily of $L\gtrsim L^*$ galaxies, and greatly increases the number of candidates for the UV brightest galaxies. Based on our population synthesis analyses, we find that average galaxies in our sample span roughly a decade in stellar mass, from ${\rm log}(M_*/M_\odot)\approx 10 - 11$, and the best-fit population synthesis ages of 200$-$400~Myr in the UV bins considered here. 

\begin{figure*}[t]
\epsscale{1.1}
\plottwo{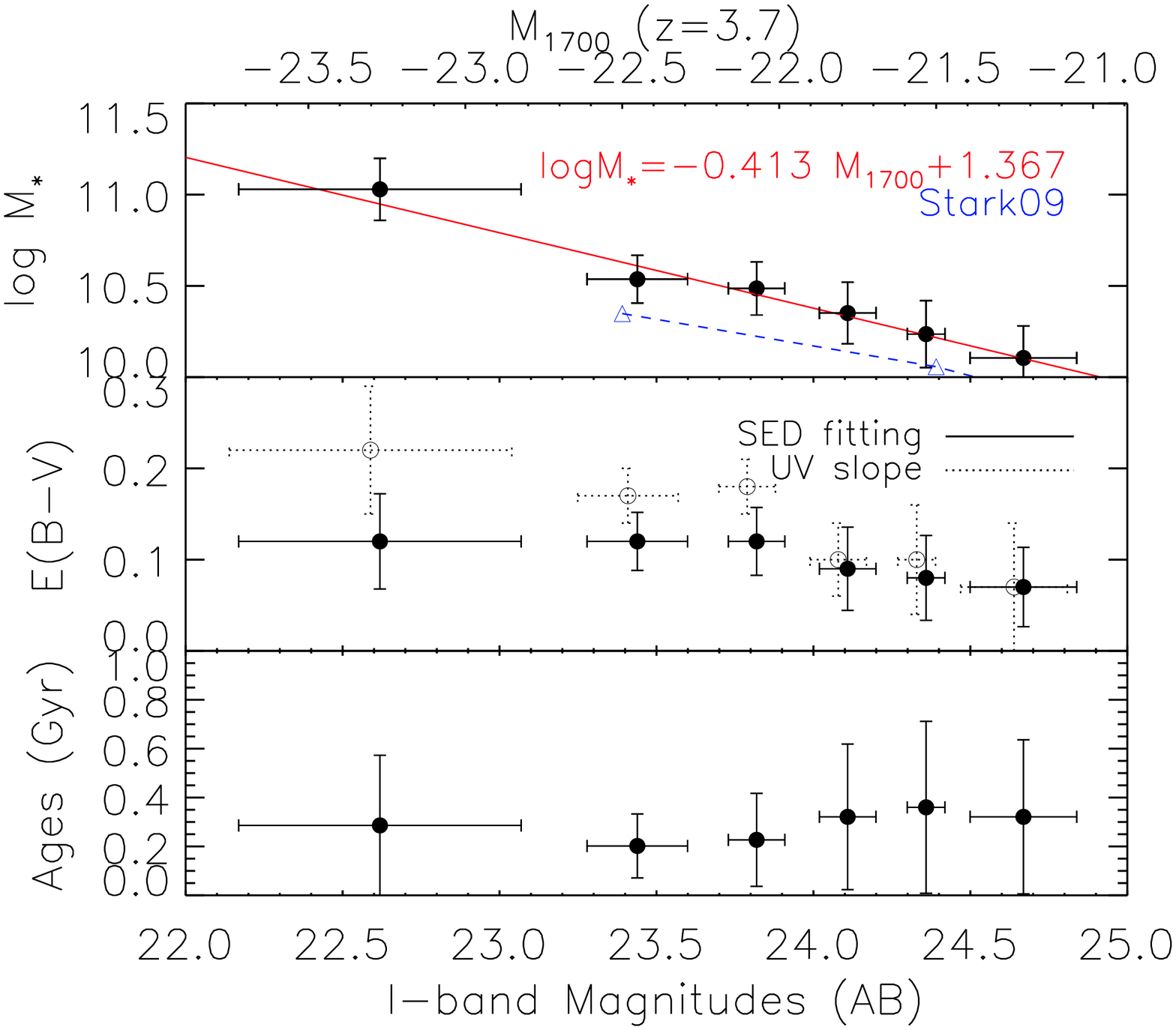}{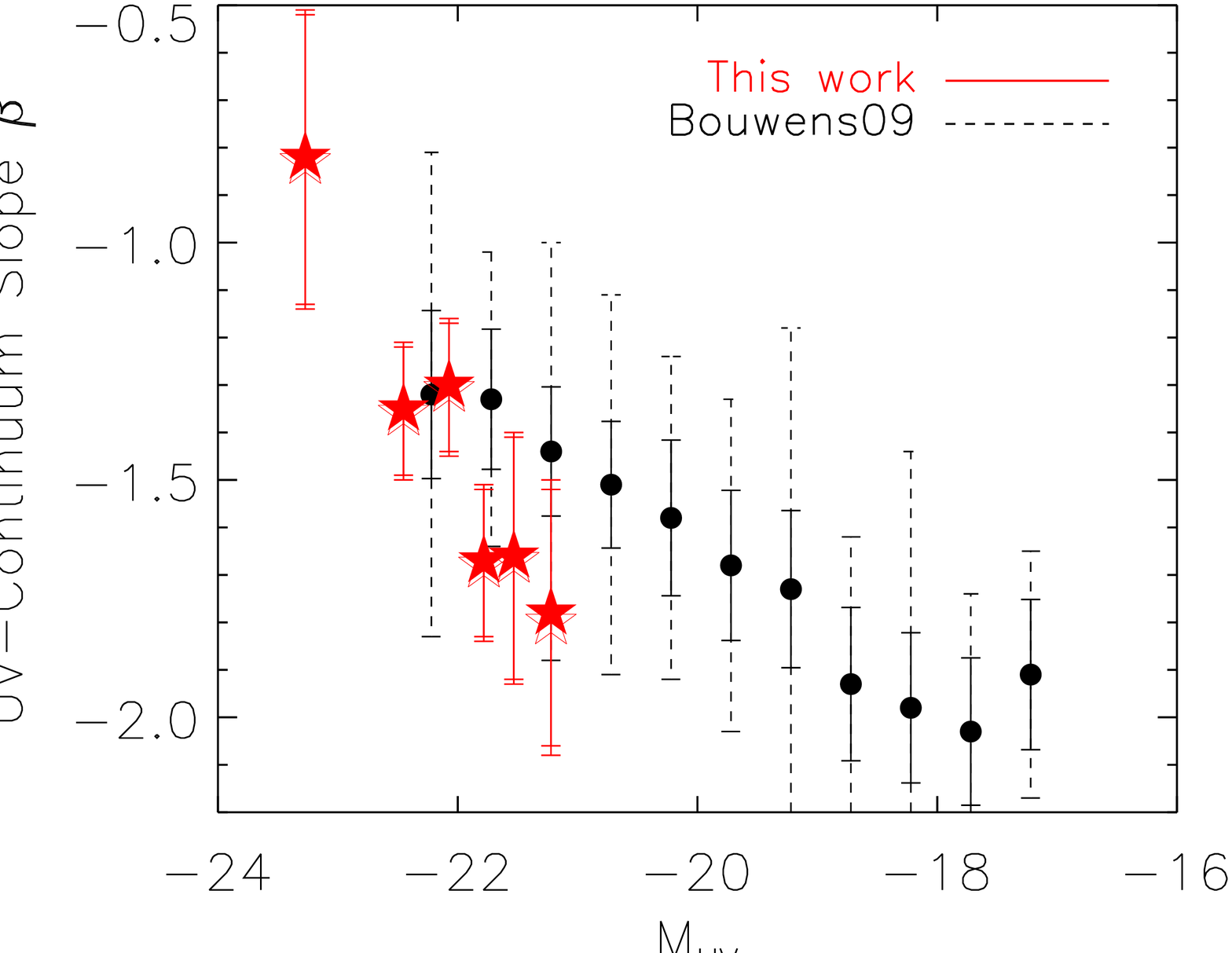}
\caption[SEDfit]{{\it Left:} The best-fit parameters for the six UV bins are shown. ({\it Top}) The mean stellar mass of galaxies rises steeply with UV luminosity (or $I$-band magnitudes) implying that galaxies with higher SFRs are also more massive. We also show the formal best-fit to our measurements (solid line) and that of \citet[][dashed line]{stark09}. ({\it Middle}) Dust reddening also increases with UV luminosity. The best-fit extinction values from the SED fitting (filled circles) and those computed directly from the UV slope (open circles; offset slightly for clarity) using the \citet{meurer99} correlation are shown. ({\it Bottom}) The inferred best-fit ages indicate young ages for all six subsamples and show little or no correlation with UV luminosity. {\it Right:} The median UV slopes $\beta$ at different UV luminosities found from this work (filled stars) are compared with those of \citet{bouwens09} shown in filled circles. For this work, the UV slopes were measured from the stacked (median) $RIJ$ photometry when the median redshift $z=3.7$ was assumed. The $RIJ$ photometry corresponds to the rest-frame wavelength range of $1300-2600$ \AA\ at $z=3.3-4.0$.  As for the \citet{bouwens09} points measured from a single UV color, both mean (solid error bars) and intrinsic scatter $1\sigma$ (dashed error bars) are shown. The two measurements are in quantitative agreement with each other, and show that the average dust extinction of galaxies increases with UV luminosity over a wide range of $M_{UV}=-23$ to -17. }
\label{sedfit}
\end{figure*}

We summarize our results in Figure \ref{sedfit}  when assuming the constant SFH model, which facilitates direct comparison with other studies in the literature. However,  we note that linearly rising SFH models generally provide equally good fit to the stacked SEDs.  The average stellar mass of our samples is tightly correlated with the UV luminosity, as expected from the homologous SED shapes in the different luminosity bins (Figure \ref{stacked_sed}). Assuming the median redshift $z=3.7$ to convert the $I$-band AB magnitude to the observed UV luminosity, we find that the correlation is well described by 
$$\log{M_*/M_\odot} = -0.413~M_{1700}+1.367$$ 
(top panel of Figure \ref{sedfit}). Since the uncertainties in stellar mass fully account for both random and intrinsic scatter in the photometry within the sample, and the width of the redshift distribution, the intrinsic correlation is likely even tighter than measured here. 
The UV luminosity range of the NDWFS sample overlaps with the luminous end of the sample (about 70 - 80 galaxies) probed by \citet{stark09} at $M_{UV} \lesssim -21.2$ at similar redshifts. Our estimation of the median stellar mass at these UV bins is slightly higher than theirs as shown in Figure \ref{sedfit} (dashed line; we adopted the stellar mass when constant star formation is assumed), but considering the large scatter and small number of galaxies included in these bins, the two measurements are in qualitative agreement with each other. We further compare our results with that of \citet{shapleyetal05}, who measured the stellar mass and UV luminosity of moderately UV-luminous star-forming galaxies at $z\sim2.3$. While they found no correlation between the UV luminosity and stellar mass in contrast to our results, their estimation of the median stellar mass is very consistent with ours, $\log{M_*/M_\odot}=10.0\pm0.2$\footnote{This is excluding five galaxies in their sample with very red UV to optical color $\mathcal{R}-K_S>3.5$. Similar sources at $z\sim3.7$ would be bright enough to be detected ($>5\sigma$) in the IRAC [3.6$\mu$m] band. Such sources are unlikely to be common in the galaxies in our sample as less than 5-10\% of our candidates are formally detected in the SDWFS data. }, at $M_{UV}\sim-21.0$, which is our faintest bin.  Further comparison is not possible at this time as only $\approx$20 galaxies in their sample are bright enough to be in ours. 

Unlike stellar mass, we find no correlation between the median age of the stellar population and the UV luminosity. The best-fit age values (200-400 Myr in all cases), suggest that the majority of  galaxies are dominated by young star-forming populations -- not just in their rest-frame UV light, but also in their rest-frame optical to near-infrared light -- that is considerably younger than the age of the Universe at $z\sim3.7$ \citep[$\approx 1.7$ Gyr: e.g.,][]{papovich01}.  Assuming exponentially declining SFHs would generally make best-fit ages even younger. The  young ages are inferred from the weak Balmer breaks between the $H$- and $K_S$-band, and are thus robust against different SFHs assumed for the SED fitting. Finally, our results show that there is a positive correlation between reddening and observed UV luminosity reflecting the increasingly steeper UV slope with luminosity observed in the stacked SEDs (Figure \ref{stacked_sed}). This extends to higher luminosity and redshift the trends observed in other Lyman break galaxy populations \citep[e.g.,][]{papovich01, bouwens09}.

\subsection{Dust Content}

We confirm the previous observations that more UV-luminous star-forming galaxies at high redshift are dustier  \citep[see, e.g.,][at $z\sim2$, 3, and $\gtrsim 4$, respectively]{reddy10, shapley01, bouwens10} as shown in Figure \ref{sedfit} (both panels). In addition to the reddening parameters derived from the population synthesis analyses, we can also directly measure the UV slope $\beta \equiv d{\rm log}F_\lambda/d{\rm log}\lambda$ from the stacked SEDs ($RIJ$ bands: rest 1300 - 2600 \AA), after correcting for the intergalactic absorption \citep{madau95} which affects the short-wavelength bands. From our UV-brightest bin to the least luminous bin, the slope changes from $\beta=-1.16\pm0.31$ (Sample \#1) to $-1.37\pm0.14$ (\#3) to $-1.78\pm0.28$ (\# 6). Shifting the median redshift by $\pm$0.2 does not significantly alter our results ($\lesssim 10$\% in the estimated UV slope $\beta$). Using the Calzetti extinction law and the correlation between the UV slope and reddening \citep{meurer99, calzetti00}, these values correspond to the reddening $E(B-V)=0.22\pm0.07, 0.17\pm0.03, 0.07\pm0.07$\footnote{$A$(1600  \AA) = $2.31(\beta-\beta_0)$ \citep[Eq. 9 in ][]{calzetti00} is used for this conversion where they adopted $\beta_0=-2.1$ for  the intrinsic UV spectral slope assuming a constant star formation history for 1 Gyr \citep{leitherer99}. The intrinsic slope does not change significantly for the same SFH for 300 Myr, the median age found for our sample.}.  The difference between the extinction found by the SED population synthesis fitting and direct measurements of the UV slope is likely due to the fact that  the stellar age is allowed to vary freely for the former but not for the latter (see footnote 15).

Our measurements are largely complementary to the UV slope determination made by \citet{bouwens09}. While they also measured the UV slope of UV-luminous ($L\gtrsim L^*$) galaxies, the majority of galaxies in their luminous samples comes from the GOODS data (0.1 deg$^2$ in total), and thus the total number of galaxies included in these bins is much smaller than ours (by a factor of $\gtrsim 10$). In Figure \ref{sedfit} ({\it right}) we show these results together. The two measurements are in quantitative agreement with each other while there are slight discrepancies at $L\approx L^*$. This may be in part due to the longer wavelength baseline used for our measurements ($RIJ$; out to $1 \mu m$) than theirs ($i'z$) where the UV slope $\beta$  steepens (see, e.g.,  Figure \ref{stacked_sed} near the rest-frame 1$\mu$m).  Furthermore, inclusion of the $R$-band in the determination of the UV continuum slope may result in, at least for some galaxies, contamination of flux from Ly$\alpha$ in emission, and as a result, cause the UV continuum to appear bluer than it is. Indeed, some ($\gtrsim$ 30\%) of our spectroscopic sample have been observed with Ly$\alpha$ in emission. However,  a much larger spectroscopic sample is needed to make statistical estimates of the fraction of galaxies with Ly$\alpha$ in emission as well as the distribution of the equivalent widths, and thus how this will affect the UV slope of each stack. Finally, we note that the selection bias of the photometric candidates also contributes to the measurement differences; for a fixed $I$-band (detection band) flux, galaxies with bluer $R-I$ colors generally have a higher completeness, and thus the median UV slope of the sample will be bluer (steeper) than the intrinsic one (see Figure 3 of \citet{bouwens09}, for example). However, correcting for this effect statistically is not straightforward in our case, as we use the stacked photometry for the $J$-band to determine the UV slope. Without knowing the intrinsic distribution of $I-J$ colors (which is the quantity we want to measure), it is difficult to statistically correct the UV slope measured from $RIJ$ photometry based on the $R-I$ colors alone. However, this explanation is consistent with the fact that the most discrepant measurements are where our selection completeness is the lowest ($\approx L^*$). 


When these effects are properly accounted for, the two measurements of the UV continuum slope are in good agreement. The overall trend is clear; the average UV continuum slope increases with UV luminosity over a wide range of luminosities ($0.02L^* - 6L^*$). \\

\section{Discussion}
We have identified the largest photometric sample to date of the UV-bright star-forming galaxies at $z\sim3.7\pm0.4$ by applying a standard color-selection technique to the broad-band data from the NDWFS. Our final sample consists of 1913 galaxies in the luminosity range of $L^*$ -- $5L^*$, distributed over an area of 5.3~deg$^2$. We have divided the sample into 6 bins of UV luminosity (i.e., observed $I$-band magnitude) and constructed the average spectral energy distributions in each bin. As described above, we confirm a strong trend between UV luminosity and stellar mass. In this section we discuss the implications of these results for the nature of the high-redshift star-forming galaxy population. 

\subsection{The Mass Dependence of the Star Formation Rate}

The fact that the more UV-luminous galaxies are considerably more massive than less UV-luminous galaxies implies a strong correlation between 
the instantaneous star-formation rates (derived from the rest-frame UV luminosity) and stellar mass (derived from the population synthesis fits, and primarily the mid-infrared photometry). Given the young mean stellar ages (200$-$400 Myr) inferred for the (luminosity weighted) stellar population in these galaxies (\S{4}), 
the tight correlation between the current forming rates of stars and existing stellar content also implies that the formation of the bulk of their stellar content is likely associated with the current episode of star formation, rather than any previous ones. 
Furthermore, we infer that the average star formation histories of super-$L^*$ galaxies are smoothly varying and long (i.e., at least comparable to the width of the redshift distribution $N(z)$, or $\gtrsim 400$ Myr) rather than peaky/bursty, as the latter would lead to little or no correlation between UV luminosity (or SFR) and stellar mass. 

A long, continuous star formation implied by our results is apparently at odds with the relatively young ages suggested by a weak Balmer break observed in all  UV bins.  Within the conventional ``$\tau$ model'', similarly young ages and the tight $SFR$-$M_*$ relation can be achieved only if the formation of galaxies began at the same point in time. It is, however, unphysical to imagine that the majority of the galaxies (of all luminosities) have such a synchronized formation redshift (at $z_f\approx 4.2-4.5$) and star formation histories!  To maintain a more physical picture, we argue that the average star formation history of the UV-bright LBG population needs to be one in which the SFR {\it increases} with time. Even models in which the SFR increase mildly with time would keep the age of the stellar population relatively young, since the light from the galaxy is always dominated by recently formed stars,  while maintaining the tight $SFR$-$M_*$ correlation. We note that similar arguments have been made in the literature to explain the same dilemma found for $z\sim2$ star-forming galaxies \citep{maraston10}, to explain the observed change in the UV luminosity and stellar mass growth with redshift \citep{papovich10}, and  to model reionization \citep{finlator10}.

\subsection{Implications for Star Formation Histories}

The observed relation between the star formation rate and stellar mass ($\Psi$-$M_*$ scaling law, hereafter) holds the direct constraint on the rate at which the SFR increases with time. 
While a more robust determination needs to be based on photometric measurements of individual galaxies with spectroscopic redshifts (to accurately measure both the mean and the scatter of the relation), we can make a rough estimate based on our stacked photometric results. For simplicity, we assume that all galaxies in our sample lie at $z=3.7$ and adopt the \citet{kennicutt98} calibration\footnote{In light of our discussion (\S 6.1) on rising star formation history, it is worth mentioning the uncertainty of this commonly used conversion on different SFHs: $\rm{SFR}(M_\odot ~{\rm year}^{-1}) = 1.4 \times 10^{-28}~ L_\nu (\rm{ergs}~ \rm{s}^{-1}~\rm{Hz}^{-1}$). For a fixed IMF, the conversion factor is lower by 10-15\% (7\%) at population age of 100 (300) Myr for star formation histories that rise linearly with time. As for exponentially declining $\tau$ models, the factor is higher by less than 5\% for $\tau\geq 300$ Myr at population ages older than 100 Myr. Considering the mild dependence of the conversion factor on population age and different SFH models, our estimation of the $\Psi$-$M_*$ scaling law should be insensitive to the range of stellar population parameters of galaxies within a given UV bin.} to convert the median $I$-band magnitude to UV SFRs for each bin, after correcting for extinction (derived from the best-fit SED model of each UV bin). The logarithmic slope of the $\Psi$-$M_*$ relation ranges over 0.8-1.1 and depends sensitively on the adopted extinction parameter. When the slope is fixed to unity, the best-fit yields the relation $\Psi(z\sim3.7)=(240\pm30~ M_\odot \rm{yr}^{-1})[M_{*}/10^{11}M_\odot]$ implying that these galaxies double their stellar mass every $290\pm40$ Myr\footnote{
The star-formation rate is the rate of increase in stellar mass, or $\Psi\equiv dM_*/dt$ (modulo mass loss). Hence, a linear relation $M_* = \tau\Psi$ implies $M_*(t) \propto e^{t/\tau}$ for galaxies to maintain the scaling relation with cosmic time. The mass-doubling time is given as $\tau \ln{2}$. 
} . These results are very similar to that found for star-forming galaxies at $z\sim2$ \citep{daddi07a} where the $\Psi$-$M_*$ relation is well described by $\Psi(z\sim2)=(200~ M_\odot \rm{yr}^{-1})(M_{*}/10^{11}M_\odot)^{0.9}$, and are somewhat higher than that found for IRAC-selected star-forming galaxies at $z\sim3$ \citep{magdis10} for which the scaling relation of $\Psi(z\sim3)=(330~ M_\odot \rm{yr}^{-1})(M_{*}/10^{11}M_\odot)^{0.91}$ is found. Our results suggest that the specific star formation rate (sSFR), on average, remains roughly constant or changes only mildly at best with stellar mass \citep[see, also,][]{daddi07a, dunne09, pannella09, magdis10, kajisawa10, reddy06a, reddy10}.

While our stacking results provide useful insight into how the sSFR changes with stellar mass, especially in the luminosity regime where no other measurements are currently available, we caution readers that the current data in hand is not adequate to constrain the full distribution of sSFR at given stellar mass bins \citep[see, e.g., ][based on individual measurements]{reddy06a,reddy10} as one needs to properly factor in the full redshift distribution and K-correction into the error budget. Furthermore, to infer the scaling law between sSFR and stellar mass, we made an additional assumption that all galaxies participate in the tight $\Psi$-$M_*$ relation with no significant outliers. In reality, it is still possible that there may be a small fraction of very massive galaxies (say, $M_*=10^{11}M_\odot$) that are not very UV-luminous (say, $I=24.5$) as the star formation started to fade. While such a population would not bias our primary measurements (i.e., the median stellar mass in a given UV bin) as long as their fraction is low, they can lower the median specific star formation rate at a fixed stellar mass.  


\subsubsection{Simple Prescriptions for the Star Formation History}

In order to develop some intuition regarding the potential of the $\Psi$-$M_*$ relation to constrain the underlying star formation history of a galaxy population, we investigated a few simple toy models in which the SFR is independent of mass, but solely a function of time, and which start forming stars at a common redshift. While such models naturally result in a constant specific star formation rate at any subsequent redshift of observation, different models will yield different normalizations at different redshifts, which can potentially be used to discriminate between models. In particular, we can compare the predictions of the toy models to the observed $\Psi$-$M_*$ relationships at $z\approx 3.7$ (this work) and at $z\approx 2$ \citep[from the work of][]{daddietal04b,daddi07a,reddy06a}. 

We begin by considering four toy models which share the common characteristic that star-formation rates are smoothly rising (or constant) with cosmic time:
\begin{enumerate}
\item  Exponentially Rising Model (EXP): \\$\Psi(t)=\Psi_0 \exp{[(t-t_0)/\tau]}$ 
\item Power-law Rising Model (POW): \\$\Psi(t)=\Psi_0 (t/t_0)^\alpha$
\item Constant Star Formation (CSF): \\$\Psi(t)=\Psi_0$
\item Delayed Star Formation (DLY): \\$\Psi(t)=\Psi_0 \exp{[-(t-t_0)/\tau] (t/t_0) }$
\end{enumerate}
The simplicity of these model parametrizations makes them an attractive starting point, before considering more complex models  \citep[see also][]{renzini09, noeske07}. All models have only one free parameter (or, in the case of the CSF model, none), which facilitates direct comparison between different models. In what follows, we give a brief description of our toy models. 

\begin{figure*}
\epsscale{1.1}
\plottwo{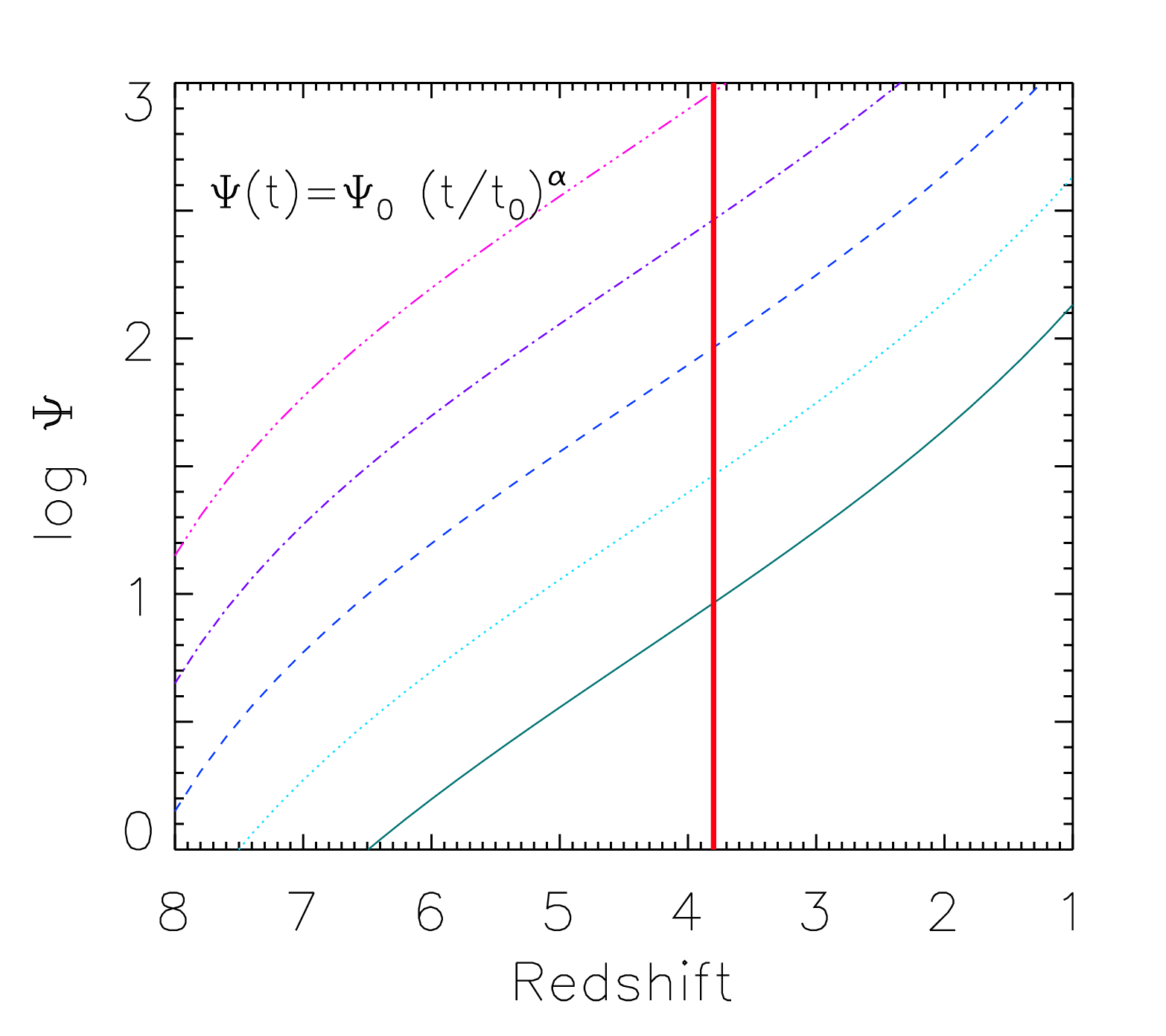}{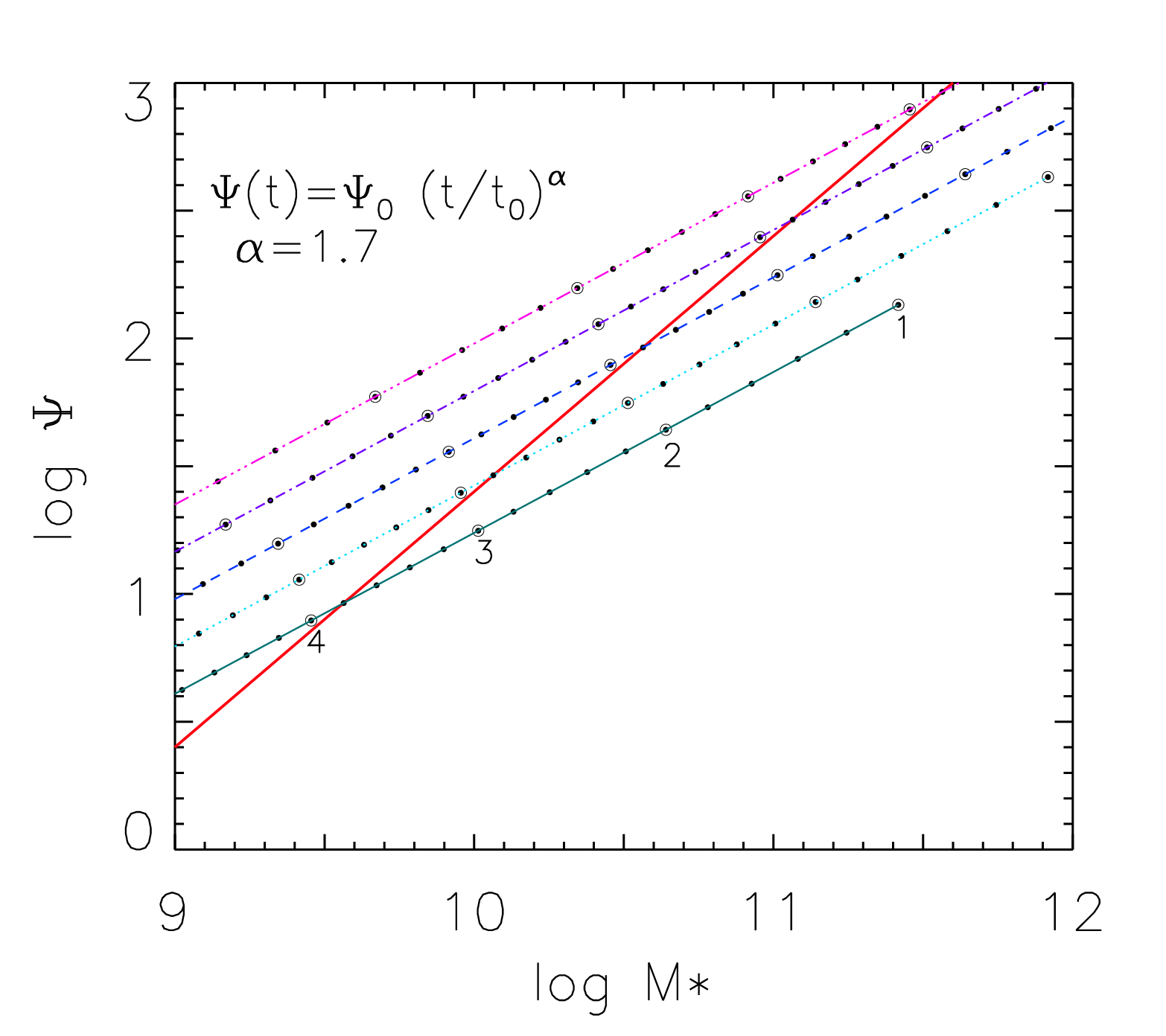}
\caption[toymodel]{{\it Left:} Star formation histories of a power-law model is shown at five different amplitudes as an example when the formation redshift $z_f=9$ and $\alpha=1.7$ are assumed. The stellar mass at redshift $z$, $M_*(z)$, to the zeroth order, is the time integral of the star formation rate $\Psi$ from the formation redshift $z_f$ to $z$. The vertical line indicates roughly where the median redshift of the $B_W$ band dropouts lies. {\it Right:} The $\Psi-M_*$ scaling relation is shown for the same SFHs. Each of the five parallel  lines indicates the time trajectory of a galaxy following the SFH shown on left. As the galaxy assembles mass, each trajectory travels from left to right in the figure with cosmic time. For the model with the lowest amplitude $\Psi$, the redshifts of various cosmic epochs are marked. The $\Psi-M_*$ scaling relation at a given cosmic epoch can be obtained by connecting different trajectories (shown for $z=3.8$ as a steeper solid line).  }
\label{toymodel}
\end{figure*}

Each galaxy is assumed to have formed at $z=z_f$ and followed a given star formation history with cosmic time. All models are normalized at $t=t_0$ to have the star formation rate $\Psi(t_0)=\Psi_0$ where time $t(z)$ is defined as the time elapsed since the formation redshift $z_f$. The location of the same galaxy on the $\Psi$-$M_*$ plane with cosmic time can be followed at each time step or $z$. Similar trajectories can be computed for other galaxies of the same SFH but with different  normalizations $\Psi_0$. The $\Psi$-$M_*$ scaling relation at a fixed redshift is then obtained by  locating all galaxies of different normalizations at the same time step $t(z)$. In Figure \ref{toymodel} we illustrate our calculation for the power-law (POW) model. The  $\Psi$-$M_*$ scaling law obtained with any smoothly rising SFH is predicted to have a logarithmic slope of unity, given our assumption of synchronized formation redshifts for all galaxies. Reproducing a shallower or steeper slope requires either the formation redshift $z_f$ or SF time scales (given as $\alpha$ or $\tau$ in our toy models) to have a mass dependence similar to that discussed by \citet{noeske07}. For example, requiring galaxies with a higher normalization $\Psi_0$ to have higher formation redshift would yield a shallower slope, which may be in qualitative agreement with the hierarchical picture of galaxy formation. We defer such modeling to future work as  the current measurements (derived from stacked broad-band data) lack the constraining power to such precision.

For the current modeling, we fix the formation redshift to $z_f=8$ and the redshifts at which galaxies are ``observed'' to $z=3.7$ and $z=2$ for comparison with the measured relationships. The choice of formation redshift is uncertain, but is unlikely to be critical to the current discussion. The age estimates from the population synthesis fitting suggest a minimum formation redshift of $z\ge 4.7$, and studies of the global star formation rate show evidence of galaxy formation at least to $z\gtrsim 7$ \citep[e.g.,][]{papovich10}. Because the elapsed time is very small at high redshifts, and since we are mainly concerned with the observed star-formation rates at $z\sim 2-3.7$, choosing $z_f=10$ instead of $z_f=8$ makes little difference to the results.  The inclusion of a range of formation redshift and more realistic redshift distribution $N(z)$ (e.g., such as that shown in Figure \ref{plot_redshift}) will indeed introduce  scatter to the observed $\Psi$-$M_*$ scaling relation. While we postpone more detailed analyses  to future work, we note that the amplitude of such scatter is likely moderate ($\lesssim 0.1$ dex). 
Given the narrow observed redshift distribution (at least for the $z\approx 3.7$ sample) and the fact that we are investigating an $\Psi$-$M_*$ relations derived from stacked data, it is unlikely that these issues will significantly affect our results.  


\subsubsection{Comparison to Observations}
\label{comparison_to_obs}

\begin{figure*}[t]
\epsscale{1.1}
\plottwo{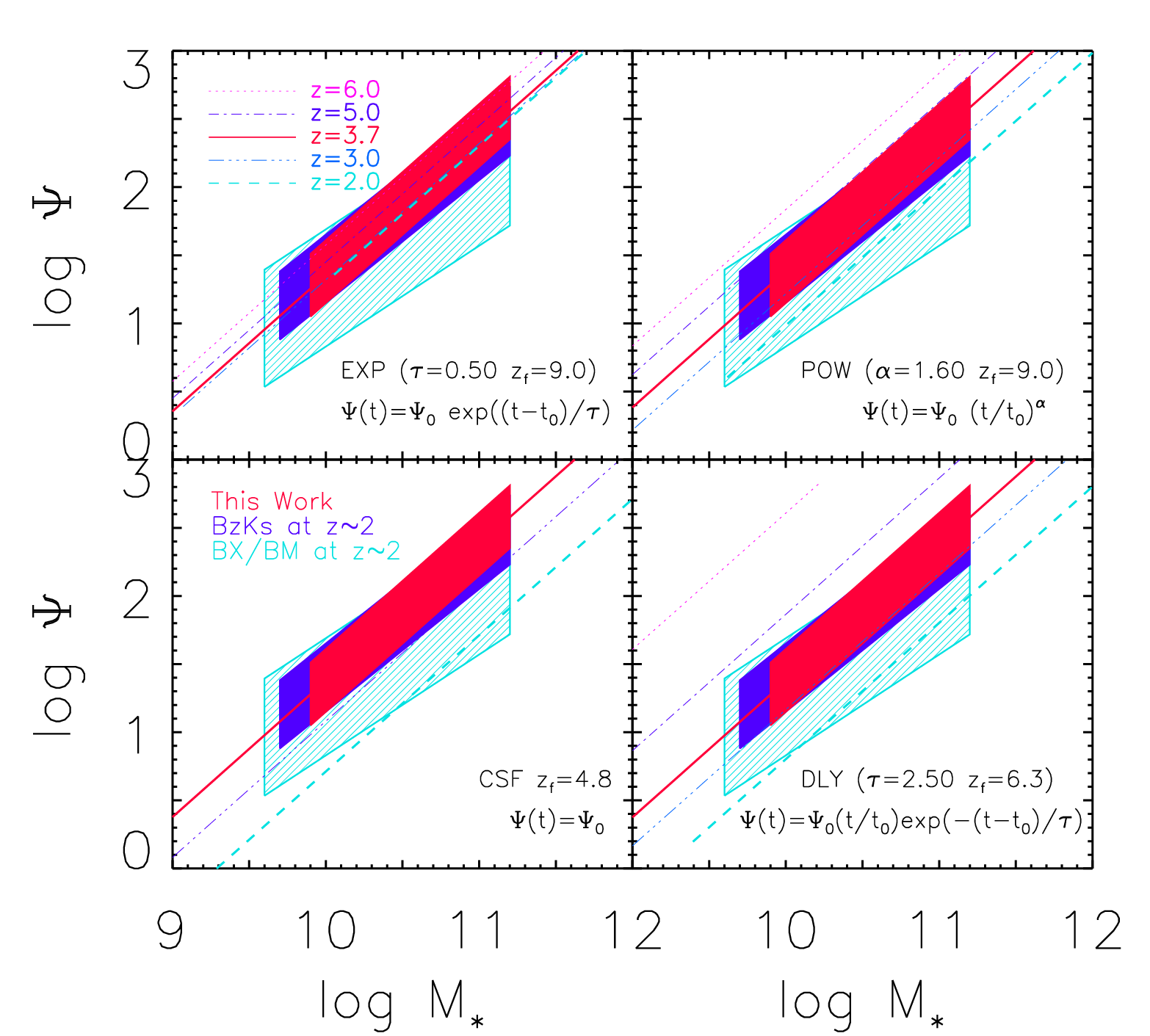}{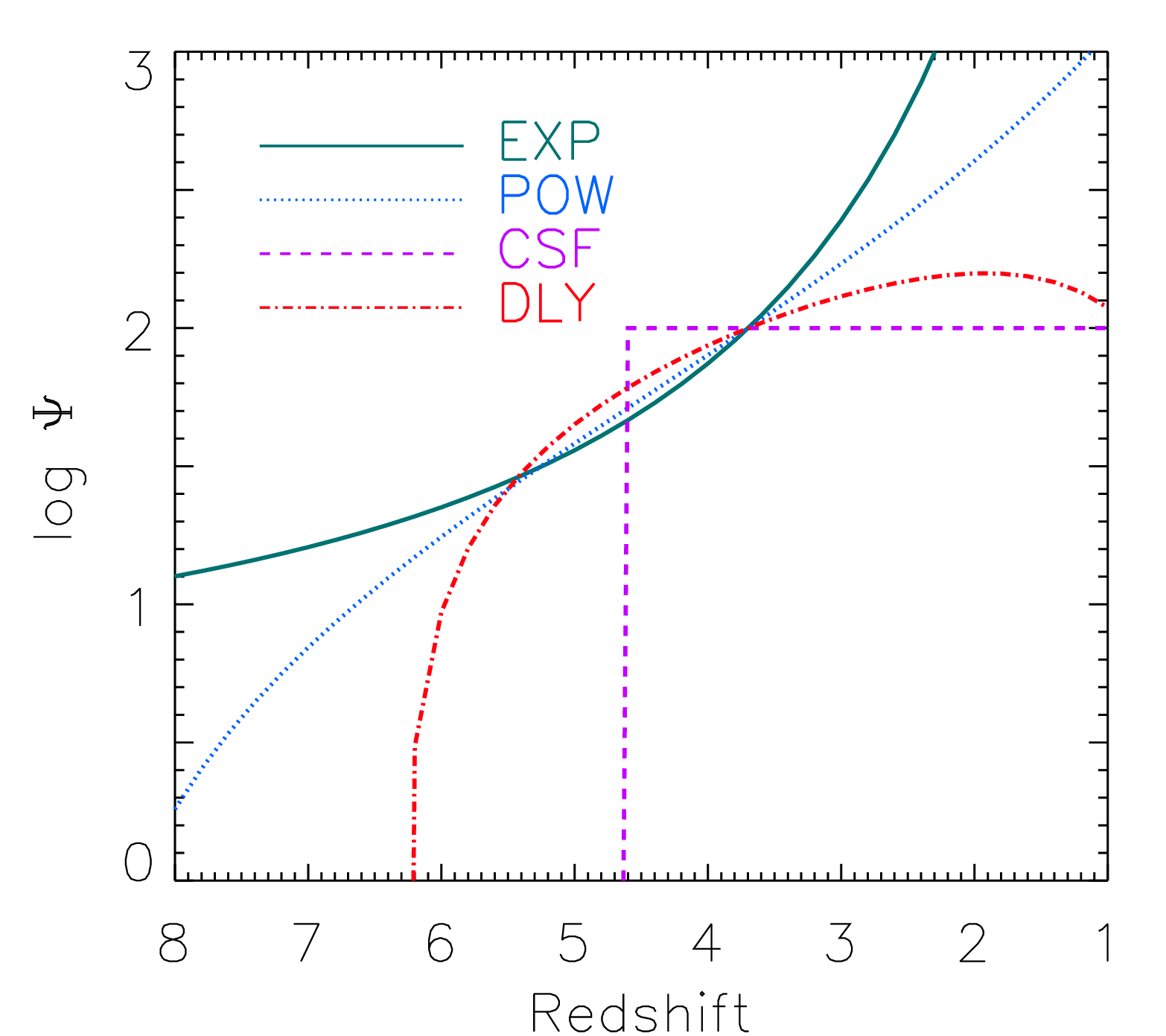}
\caption[SEDfit]{{\it Left:} The redshift evolution of the $\Psi$-$M_*$ relation is explored for four SFH models from $z=6$ to 2 in comparison with  the observations at $z\sim3.7$ (this work) and $z\sim2$ \citep{daddi07a, reddy06a}. The parameters $\alpha$ or $\tau$ and the formation redshift $z_f$ are chosen for each model to best reproduce the observed relation at $z\sim3.7$. The EXP and POW models with a suitable choice of the model parameter may provide the reasonable description of the existing data while the CSF and DLY model do not as they predict a steep decline in the scaling relation with cosmic time (see text). {\it Right:} Four SFH models  that best-fit the observed $\Psi-M_*$ scaling relation at $z\sim3.7$ are shown. All models are normalized to have $\Psi=100~ M_\odot \rm{yr}^{-1}$ at $z=3.7$. The time evolution of the SFH (before and after the time of observations)  and, as a result, the $\Psi-M_*$ scaling law vary dramatically for different models. Hence, more robust determination of the observed $\Psi-M_*$ scaling relations at different redshifts has the potential to elucidate the star formation histories of high redshift galaxy populations and their descendants at later epochs. }
\label{sfh_models}
\end{figure*}

Following this simple procedure, we investigate how each model fares against the observational constraints. First, we compute the $\Psi$-$M_*$ scaling relation at $z\sim 3.7$ and determine the best-fit parameter (given as $\alpha$ or $\tau$) by matching the model predictions to our measurements (see Figure \ref{sfh_models}a for the time evolution predicted by each model). Then using the best-fit parameter, we compute the scaling relation at redshift 2 to 6 to follow the redshift evolution of each model. As the observational constraints (of individual measurements) at other cosmic epochs are limited at the present time, we mainly focus on the measurements at $z\sim3.7$ (this work), $BzK$ galaxies at $z\sim2$ \citep{daddi07a, pannella09}, and BX/BM star-forming galaxies at $z\sim2$ \citep{reddy06a} to compare with our models. We note that the scaling relation from individual measurements by \citet{daddi07a} is similar to that measured from the stacked radio data \citep{pannella09} of the same $BzK$ sample. In Figure \ref{sfh_models}b we show the redshift evolution of all four models and best-fit model parameters together with the observations at $z\sim3.7$ and $z\sim2$ (shaded regions). The $1\sigma$ scatter of 0.25 dex is added for the $B_W$ dropouts to account for the redshift distribution and measurement errors while for $BzK$s we use the published value of 0.23 dex\footnote{The quoted value of the semi-interquartile range of 0.16 dex \citep{daddi07a} corresponds to 0.23 dex in the standard deviation assuming normal distribution. } \citep{daddi07a}. As for BX/BM galaxies, the scatter is observed to be $\approx$ 0.46 dex, about 70\% larger than that for $BzK$s.

It is evident from Figure \ref{sfh_models}b that the CSF and DLY model predict $\approx 0.5$ and $0.7$ dex decline in the amplitude of the $\Psi$-$M_*$ scaling relation, clearly too steep to be supported by the observations at $z\sim 2$ and $3.7$. In particular, the CSF model requires that the formation redshift is $z_f \approx 4.8$ ($\approx$500 Myr prior to the observations) to reproduce the observed relation at $z\sim 3.7$\footnote{Note that the {\it exponentially  declining} $\tau$ models will generally behave in a similar way to CSF models. However, for a given $[\Psi, M_*]$ value, they require the formation redshift to be even closer to the time of observations than CSF models because the SFR was higher in the past.} The implication of the relatively low formation redshift ($z_f \approx 4.8$ for the $B_W$-band dropouts, and $z_f \approx 2.3$ for $z\sim2$ galaxies) is that the galaxy samples observed at different redshift bins cannot be connected to one another through cosmic time. Such a scenario would not only contradict long duty cycles suggested by the tight $\Psi$-$M_*$ scaling relation but also would require an unrealistic picture in which the majority of the galaxies observed at $z\sim3.7$ would suddenly start forming stars at a very short time interval around $z=4.8$.  

The EXP model provides a good description of the observations at both redshifts when $\tau=0.50$ Gyr is assumed (consistent with the age constraints). A slight change of the formation redshift ($z_f=7$ or $11$ instead of $8$) does not change the timescale more than 50 Myr. The POW model also provides a good description at $z\sim 3.7$ when $\alpha=1.60$ (1.70) and $z_f=9$ (11) is assumed.  It is interesting that the exponent $\alpha$ found by simply requiring it to reproduce the scaling relation at $z\sim3.7$ is very similar to that found by \citet{papovich10} who obtained the exponent $\alpha$ by fiting the redshift evolution of galaxy samples at a fixed number density at $z=3-8$ (see their Figure 2). Unlike the EXP model, the POW model predicts a moderate evolution in the amplitude of the scaling relation ($\approx 0.3$ dex from $z=3.7$ to $2$). While the predicted decline is formally not consistent with the $BzK$ measurements, the real uncertainties (random and systematic) are likely under-represented in Figure \ref{sfh_models}b ({\it shaded area}). We note that \citet{reddy06a} reported much larger scatter ($\approx 0.4-0.5$ dex) for BX/BM star-forming galaxies at the similar redshift range. Furthermore, there are other studies that have found similarly large or even larger scatter at $z\sim3$ and a higher normalization than the $B_W$ dropouts in the scaling law \citep{mannucci09, magdis10}. Such discrepancies in the literature highlight the existing uncertainties and difficulties in measuring such quantities at high redshift. 

In summary, our analyses suggest that both exponentially rising or power-law rising models do better than constant SFH model. A more definitive answer will have to await better data on samples at multiple redshifts. The existing data clearly suffers from different selection biases, the random (and likely significant) photometric errors in the rest-optical photometry of high-redshift galaxies, and the uncertainties associated with SED modeling assumptions. Nevertheless, it is worth noting that comparing the observed $\Psi$-$M_*$ relations at different redshifts to simple models, such as those discussed here, has the potential to elucidate the star formation histories of high redshift galaxy populations. The key to discriminating between, say, SFHs that rise exponentially or as a power-law lies in unambiguously determining the overall amplitude of the $\Psi$-$M_*$ scaling relation at multiple redshifts. While interesting, the prescriptions for star formation histories described above are very simplistic and one can easily imagine that SFHs of galaxies can be mass dependent. However, considering more detailed models is beyond the scope of the current work, and unwarranted by the data. A future 
determination of the scatter in the relationship (through measurements of individual galaxies, as opposed to stacked samples) will bear directly on the distribution of star formation history parameters (e.g., formation redshifts, star formation rate amplitudes, or SFH modes) for the galaxy population. \\

\subsection{What is the Duty Cycle of Star Formation?}

The UV luminosity of a star-forming galaxy is dominated by hot stars and therefore measures the instantaneous star formation rate of the system. However, the existence of the tight $\Psi$-$M_*$ relation observed within the $z\approx3.7\pm0.4$ star-forming population suggests that the duty cycle of star formation must be significantly longer than the time range sampled, i.e., $\gtrsim 400$~Myr. 
Furthermore, the positive correlation between the two quantities --  $dM_*/dt \propto M_*$ -- yields the solution of $M_*\propto t^\alpha$ or $\propto \exp{[t/\tau]}$ where $\alpha$ or $\tau$ is positive, unless the galaxies are observed in a special time of their formation histories. If the onset of star formation in these galaxies is not coordinated, the direct implication would be that the star formation rate is increasing with time. The rate of increase in star-formation rates would be better determined by observing similar relations at different cosmic epochs as discussed in \S \ref{comparison_to_obs}. 
 In such a scenario, most galaxies are expected to have a fairly long duty cycle until the process is terminated or slowed down by a feedback mechanism.

A complementary constraint on the duty cycle of star formation can be obtained from clustering measurements: the measured clustering strength (combined with a theoretical model) can provide an estimate of the dark halo mass hosting the typical sample galaxy, and a comparison between the expected number density of these halos and the observed number density of the sample galaxies yields an estimate of the duty cycle \citep[e.g.,][]{martini01,lee06,lee09,mclure09}. 
Indeed, the long duty cycle suggested by our results is supported by the clustering studies of luminous high-redshift galaxies \citep[e.g.,][]{adelberger05, ouchi05} and the interpretations in relation to the theoretical predictions of dark matter halo clustering at the same epoch \citep{conroy06}.


In contrast, for much less luminous star-forming galaxies at similar redshifts, the evidence seems to point to the contrary. \citet{lee09} found that the small-scale clustering of the faint UV-selected galaxies at $z\sim4$ and 5 (the deepest data with which the clustering of high-redshift galaxies was measured thus far) is much weaker than that for dark matter halos, and argued that it may be a result of short duty cycles ($\approx 10-30$\% or $\leq 0.4$ Gyr at $1\sigma$ level) governing their star formation. Furthermore, \citet{stark09}, who studied the SED properties of similar samples of UV-faint star-forming galaxies, found little evolution in ages for galaxies at $z\sim4$, 5, and 6. They argued that their star formation may not be continuous, at least at the constant level, as it would lead to older ages, on average, for galaxies at $z\sim4$ compared to $z\sim6$. Alternatively, as they argued, their young ages could also be produced by rising SFHs in the same way as more UV luminous galaxies.  However, such a scenario would be inconsistent with the \citet{lee09} clustering study as the small-scale clustering of the galaxies would closely reflect that of dark halos. 

Taken at face value, this apparent difference between the implied duty cycles for star formation in UV-luminous and UV-faint galaxies suggests that star-formation is a luminosity-dependent process such that luminous galaxies rapidly grow in mass by continuously forming stars (at an increasing rate) while faint galaxies flicker with cosmic time as their UV light turns on and off at short intervals ($\approx 200$-300 Myr). However, it is also true that our understanding of UV-faint high-redshift galaxies is more limited than that of more luminous counterparts as it is more challenging to obtain spectroscopic observations even to identify their redshifts, not to mention more detailed physical properties. Further investigations are needed to determine whether such differences are due to the observational limitation or an intrinsic one. \\


\subsection{The Descendants of the most UV Luminous Galaxies at $z\approx3.7$}

The extinction-corrected star-formation rates of $L\gtrsim L^*$ galaxies span 50 - 250 $M_\odot \rm{yr}^{-1}$ at the epoch of the observation (i.e., $z\approx3.7$).  The observed $\Psi$-$M_*$ relation suggests that the star-formation rates will increase with time (as $\propto e^{t/\tau}$ or  $\propto t^{1.6-1.7}$; see Figure \ref{sfh_models}a).  Assuming an exponential rise, we estimate that these galaxies will double their stellar mass every $\approx 300$ Myr until (or unless) the star formation is quenched or slowed by a feedback mechanism. In the absence of quenching, by $z=2$ (i.e., $\approx 1.5$ Gyr after $z=3.7$), $L\approx L^*$ galaxies with $\langle M_* \rangle \approx 10^{10}M_\odot$ will grow into $M_*\approx 10^{11.4}M_\odot$ while 2.5$L^*$ galaxies will become in excess of $10^{12}M_\odot$. Hence, the descendants of these UV-luminous star-forming systems at $z\approx3.7$ will likely be galaxies at the top end of the mass function by $z\approx2$. 

It is not clear if such growth in the SFR is sustainable either by the fuel supply within a galaxy or by the physical mechanisms of star formation. More detailed studies of individual star-forming galaxies can shed light on the gas content and fueling of star formation and the role of feedback processes (e.g., AGN). In addition, comparative statistical studies of the UV luminosity function, mass function, and large scale clustering of galaxies at different redshifts (e.g., $z\approx3.7$ and 2) can constrain the duty cycles of star formation and thereby determine what fraction of the UV-brightest galaxies at high redshift will have evolved into the most massive galaxies at lower redshift (Lee et al., in prep). We intend to investigate this UV-bright sample using both detailed and statistical approaches with the goal of elucidating their evolution and determining how and when the feedback processes (such as AGN feedback) will shape the future fate of these rare galaxies \citep[e.g.,][]{marchesini10}.

\acknowledgments
KSL gratefully acknowledges the generous support of Gilbert and Jaylee Mead for their namesake fellowship in the Yale Center for Astronomy and Astrophysics. 
KSL also thanks Adam Muzzin, C.~Meg Urry, and Casey Papovich for useful discussions and suggestions. 
AHG acknowledges support for this work from the National Science Foundation under grant AST-00708490. 
The analysis in this paper is partly made with the Keck observations awarded through Yale University.
This work is based in part on observations made with the {\it Spitzer Space Telescope}, which is operated by the Jet Propulsion Laboratory, California Institute of Technology. 
We are grateful to the expert assistance of the staff of Kitt Peak National Observatory where the optical and near-infrared observations of the NDWFS Bo\"otes Field were obtained. The authors thank NOAO for supporting the NOAO Deep Wide-Field Survey. The research activities of AD and BTJ are supported by NOAO, which is operated by the Association of Universities for Research in Astronomy (AURA) under a cooperative agreement with the National Science Foundation. 
 The LBT is an international collaboration among institutions in the United States, Italy and Germany. LBT Corporation partners are: The University of Arizona on behalf of the Arizona university system; Instituto Nazionale di Astrofisica, Italy; LBT Beteiligungsgesellschaft, Germany, representing the Max-Planck Society, the Astrophysical Institute Potsdam, and Heidelberg University; The Ohio State University, and The Research Corporation, on behalf of The University of Notre Dame, University of Minnesota and University of Virginia.
Some of the data presented herein were obtained at the W.M. Keck Observatory, which is operated as a scientific partnership among the California Institute of Technology, the University of California and the National Aeronautics and Space Administration. The Observatory was made possible by the generous financial support of the W.M. Keck Foundation. We thank the staffs of the W.~M.~Keck Observatory, in particular, Gregory Wirth.
The analysis pipeline used to reduce the DEIMOS data was developed at UC Berkeley with support from NSF grant AST-0071048.  The authors also wish to recognize and acknowledge the very significant cultural role and reverence that the summit of Mauna Kea has always had within the indigenous Hawaiian community.

\bibliographystyle{/Users/kyoungsoolee/publications/apj}
\bibliography{/Users/kyoungsoolee/publications/apj-jour,/Users/kyoungsoolee/publications/myrefs}  

\begin{thebibliography}{65}
\expandafter\ifx\csname natexlab\endcsname\relax\def\natexlab#1{#1}\fi

\bibitem[{{Adelberger} {et~al.}(2005){Adelberger}, {Steidel}, {Pettini},
  {Shapley}, {Reddy}, \& {Erb}}]{adelberger05}
{Adelberger}, K.~L., {Steidel}, C.~C., {Pettini}, M., {Shapley}, A.~E.,
  {Reddy}, N.~A., \& {Erb}, D.~K. 2005, \apj, 619, 697

\bibitem[{{Ashby} {et~al.}(2009){Ashby}, {Stern}, {Brodwin}, {Griffith},
  {Eisenhardt}, {Koz{\l}owski}, {Kochanek}, {Bock}, {Borys}, {Brand}, {Brown},
  {Cool}, {Cooray}, {Croft}, {Dey}, {Eisenstein}, {Gonzalez}, {Gorjian},
  {Grogin}, {Ivison}, {Jacob}, {Jannuzi}, {Mainzer}, {Moustakas},
  {R{\"o}ttgering}, {Seymour}, {Smith}, {Stanford}, {Stauffer}, {Sullivan},
  {van Breugel}, {Willner}, \& {Wright}}]{ashby09}
{Ashby}, M.~L.~N., {Stern}, D., {Brodwin}, M., {Griffith}, R., {Eisenhardt},
  P., {Koz{\l}owski}, S., {Kochanek}, C.~S., {Bock}, J.~J., {Borys}, C.,
  {Brand}, K., {Brown}, M.~J.~I., {Cool}, R., {Cooray}, A., {Croft}, S., {Dey},
  A., {Eisenstein}, D., {Gonzalez}, A.~H., {Gorjian}, V., {Grogin}, N.~A.,
  {Ivison}, R.~J., {Jacob}, J., {Jannuzi}, B.~T., {Mainzer}, A., {Moustakas},
  L.~A., {R{\"o}ttgering}, H.~J.~A., {Seymour}, N., {Smith}, H.~A., {Stanford},
  S.~A., {Stauffer}, J.~R., {Sullivan}, I., {van Breugel}, W., {Willner},
  S.~P., \& {Wright}, E.~L. 2009, \apj, 701, 428

\bibitem[{{Bertin} \& {Arnouts}(1996)}]{bertina96}
{Bertin}, E. \& {Arnouts}, S. 1996, \aaps, 117, 393

\bibitem[{{Bouwens} {et~al.}(2004){Bouwens}, {Illingworth}, {Blakeslee},
  {Broadhurst}, \& {Franx}}]{bouwens04c}
{Bouwens}, R.~J., {Illingworth}, G.~D., {Blakeslee}, J.~P., {Broadhurst},
  T.~J., \& {Franx}, M. 2004, \apjl, 611, L1

\bibitem[{{Bouwens} {et~al.}(2009){Bouwens}, {Illingworth}, {Franx}, {Chary},
  {Meurer}, {Conselice}, {Ford}, {Giavalisco}, \& {van Dokkum}}]{bouwens09}
{Bouwens}, R.~J., {Illingworth}, G.~D., {Franx}, M., {Chary}, R., {Meurer},
  G.~R., {Conselice}, C.~J., {Ford}, H., {Giavalisco}, M., \& {van Dokkum}, P.
  2009, \apj, 705, 936

\bibitem[{{Bouwens} {et~al.}(2007){Bouwens}, {Illingworth}, {Franx}, \&
  {Ford}}]{bouwens07}
{Bouwens}, R.~J., {Illingworth}, G.~D., {Franx}, M., \& {Ford}, H. 2007, \apj,
  670, 928

\bibitem[{{Bouwens} {et~al.}(2010){Bouwens}, {Illingworth}, {Oesch}, {Trenti},
  {Stiavelli}, {Carollo}, {Franx}, {van Dokkum}, {Labb{\'e}}, \&
  {Magee}}]{bouwens10}
{Bouwens}, R.~J., {Illingworth}, G.~D., {Oesch}, P.~A., {Trenti}, M.,
  {Stiavelli}, M., {Carollo}, C.~M., {Franx}, M., {van Dokkum}, P.~G.,
  {Labb{\'e}}, I., \& {Magee}, D. 2010, \apjl, 708, L69

\bibitem[{{Brown} {et~al.}(2007){Brown}, {Dey}, {Jannuzi}, {Brand}, {Benson},
  {Brodwin}, {Croton}, \& {Eisenhardt}}]{brown07}
{Brown}, M.~J.~I., {Dey}, A., {Jannuzi}, B.~T., {Brand}, K., {Benson}, A.~J.,
  {Brodwin}, M., {Croton}, D.~J., \& {Eisenhardt}, P.~R. 2007, \apj, 654, 858

\bibitem[{{Bruzual} \& {Charlot}(2003)}]{bc03}
{Bruzual}, G. \& {Charlot}, S. 2003, \mnras, 344, 1000

\bibitem[{{Calzetti} {et~al.}(2000){Calzetti}, {Armus}, {Bohlin}, {Kinney},
  {Koornneef}, \& {Storchi-Bergmann}}]{calzetti00}
{Calzetti}, D., {Armus}, L., {Bohlin}, R.~C., {Kinney}, A.~L., {Koornneef}, J.,
  \& {Storchi-Bergmann}, T. 2000, \apj, 533, 682

\bibitem[{{Conroy} {et~al.}(2006){Conroy}, {Wechsler}, \&
  {Kravtsov}}]{conroy06}
{Conroy}, C., {Wechsler}, R.~H., \& {Kravtsov}, A.~V. 2006, \apj, 647, 201

\bibitem[{{Daddi} {et~al.}(2004){Daddi}, {Cimatti}, {Renzini}, {Fontana},
  {Mignoli}, {Pozzetti}, {Tozzi}, \& {Zamorani}}]{daddietal04b}
{Daddi}, E., {Cimatti}, A., {Renzini}, A., {Fontana}, A., {Mignoli}, M.,
  {Pozzetti}, L., {Tozzi}, P., \& {Zamorani}, G. 2004, \apj, 617, 746

\bibitem[{{Daddi} {et~al.}(2007){Daddi}, {Dickinson}, {Morrison}, {Chary},
  {Cimatti}, {Elbaz}, {Frayer}, {Renzini}, {Pope}, {Alexander}, {Bauer},
  {Giavalisco}, {Huynh}, {Kurk}, \& {Mignoli}}]{daddi07a}
{Daddi}, E., {Dickinson}, M., {Morrison}, G., {Chary}, R., {Cimatti}, A.,
  {Elbaz}, D., {Frayer}, D., {Renzini}, A., {Pope}, A., {Alexander}, D.~M.,
  {Bauer}, F.~E., {Giavalisco}, M., {Huynh}, M., {Kurk}, J., \& {Mignoli}, M.
  2007, \apj, 670, 156

\bibitem[{{Dav{\'e}}(2008)}]{dave08}
{Dav{\'e}}, R. 2008, \mnras, 385, 147

\bibitem[{{Dekel} {et~al.}(2009){Dekel}, {Sari}, \& {Ceverino}}]{dekel09}
{Dekel}, A., {Sari}, R., \& {Ceverino}, D. 2009, \apj, 703, 785

\bibitem[{{Desai} {et~al.}(2008){Desai}, {Soifer}, {Dey}, {Jannuzi}, {Le
  Floc'h}, {Bian}, {Brand}, {Brown}, {Armus}, {Weedman}, {Cool}, {Stern}, \&
  {Brodwin}}]{desai08}
{Desai}, V., {Soifer}, B.~T., {Dey}, A., {Jannuzi}, B.~T., {Le Floc'h}, E.,
  {Bian}, C., {Brand}, K., {Brown}, M.~J.~I., {Armus}, L., {Weedman}, D.~W.,
  {Cool}, R., {Stern}, D., \& {Brodwin}, M. 2008, \apj, 679, 1204

\bibitem[{{Dunne} {et~al.}(2009){Dunne}, {Ivison}, {Maddox}, {Cirasuolo},
  {Mortier}, {Foucaud}, {Ibar}, {Almaini}, {Simpson}, \& {McLure}}]{dunne09}
{Dunne}, L., {Ivison}, R.~J., {Maddox}, S., {Cirasuolo}, M., {Mortier}, A.~M.,
  {Foucaud}, S., {Ibar}, E., {Almaini}, O., {Simpson}, C., \& {McLure}, R.
  2009, \mnras, 394, 3

\bibitem[{{Elbaz} {et~al.}(2007){Elbaz}, {Daddi}, {Le Borgne}, {Dickinson},
  {Alexander}, {Chary}, {Starck}, {Brandt}, {Kitzbichler}, {MacDonald},
  {Nonino}, {Popesso}, {Stern}, \& {Vanzella}}]{elbaz07}
{Elbaz}, D., {Daddi}, E., {Le Borgne}, D., {Dickinson}, M., {Alexander}, D.~M.,
  {Chary}, R., {Starck}, J., {Brandt}, W.~N., {Kitzbichler}, M., {MacDonald},
  E., {Nonino}, M., {Popesso}, P., {Stern}, D., \& {Vanzella}, E. 2007, \aap,
  468, 33

\bibitem[{{Elsner} {et~al.}(2008){Elsner}, {Feulner}, \& {Hopp}}]{elsner08}
{Elsner}, F., {Feulner}, G., \& {Hopp}, U. 2008, \aap, 477, 503

\bibitem[{{Erb} {et~al.}(2006){Erb}, {Steidel}, {Shapley}, {Pettini}, {Reddy},
  \& {Adelberger}}]{erbetal06b}
{Erb}, D.~K., {Steidel}, C.~C., {Shapley}, A.~E., {Pettini}, M., {Reddy},
  N.~A., \& {Adelberger}, K.~L. 2006, \apj, 646, 107

\bibitem[{{Faber} {et~al.}(2003){Faber}, {Phillips}, {Kibrick}, {Alcott},
  {Allen}, {Burrous}, {Cantrall}, {Clarke}, {Coil}, {Cowley}, {Davis}, {Deich},
  {Dietsch}, {Gilmore}, {Harper}, {Hilyard}, {Lewis}, {McVeigh}, {Newman},
  {Osborne}, {Schiavon}, {Stover}, {Tucker}, {Wallace}, {Wei}, {Wirth}, \&
  {Wright}}]{deimos_ref}
{Faber}, S.~M., {Phillips}, A.~C., {Kibrick}, R.~I., {Alcott}, B., {Allen},
  S.~L., {Burrous}, J., {Cantrall}, T., {Clarke}, D., {Coil}, A.~L., {Cowley},
  D.~J., {Davis}, M., {Deich}, W.~T.~S., {Dietsch}, K., {Gilmore}, D.~K.,
  {Harper}, C.~A., {Hilyard}, D.~F., {Lewis}, J.~P., {McVeigh}, M., {Newman},
  J., {Osborne}, J., {Schiavon}, R., {Stover}, R.~J., {Tucker}, D., {Wallace},
  V., {Wei}, M., {Wirth}, G., \& {Wright}, C.~A. 2003, in Society of
  Photo-Optical Instrumentation Engineers (SPIE) Conference Series, Vol. 4841,
  Society of Photo-Optical Instrumentation Engineers (SPIE) Conference Series,
  ed. {M.~Iye \& A.~F.~M.~Moorwood}, 1657--1669

\bibitem[{{Ferguson} {et~al.}(2004){Ferguson}, {Dickinson}, {Giavalisco},
  {Kretchmer}, {Ravindranath}, {Idzi}, {Taylor}, {Conselice}, {Fall},
  {Gardner}, {Livio}, {Madau}, {Moustakas}, {Papovich}, {Somerville},
  {Spinrad}, \& {Stern}}]{ferguson04}
{Ferguson}, H.~C., {Dickinson}, M., {Giavalisco}, M., {Kretchmer}, C.,
  {Ravindranath}, S., {Idzi}, R., {Taylor}, E., {Conselice}, C.~J., {Fall},
  S.~M., {Gardner}, J.~P., {Livio}, M., {Madau}, P., {Moustakas}, L.~A.,
  {Papovich}, C.~M., {Somerville}, R.~S., {Spinrad}, H., \& {Stern}, D. 2004,
  \apjl, 600, L107

\bibitem[{{Finlator} {et~al.}(2010){Finlator}, {Oppenheimer}, \&
  {Dav{\'e}}}]{finlator10}
{Finlator}, K., {Oppenheimer}, B.~D., \& {Dav{\'e}}, R. 2010, ArXiv e-prints

\bibitem[{{Genzel} {et~al.}(2006){Genzel}, {Tacconi}, {Eisenhauer},
  {F{\"o}rster Schreiber}, {Cimatti}, {Daddi}, {Bouch{\'e}}, {Davies},
  {Lehnert}, {Lutz}, {Nesvadba}, {Verma}, {Abuter}, {Shapiro}, {Sternberg},
  {Renzini}, {Kong}, {Arimoto}, \& {Mignoli}}]{genzel06}
{Genzel}, R., {Tacconi}, L.~J., {Eisenhauer}, F., {F{\"o}rster Schreiber},
  N.~M., {Cimatti}, A., {Daddi}, E., {Bouch{\'e}}, N., {Davies}, R., {Lehnert},
  M.~D., {Lutz}, D., {Nesvadba}, N., {Verma}, A., {Abuter}, R., {Shapiro}, K.,
  {Sternberg}, A., {Renzini}, A., {Kong}, X., {Arimoto}, N., \& {Mignoli}, M.
  2006, \nat, 442, 786

\bibitem[{{Giavalisco} {et~al.}(2004){Giavalisco}, {Dickinson}, {Ferguson},
  {Ravindranath}, {Kretchmer}, {Moustakas}, {Madau}, {Fall}, {Gardner},
  {Livio}, {Papovich}, {Renzini}, {Spinrad}, {Stern}, \& {Riess}}]{mauro04b}
{Giavalisco}, M., {Dickinson}, M., {Ferguson}, H.~C., {Ravindranath}, S.,
  {Kretchmer}, C., {Moustakas}, L.~A., {Madau}, P., {Fall}, S.~M., {Gardner},
  J.~P., {Livio}, M., {Papovich}, C., {Renzini}, A., {Spinrad}, H., {Stern},
  D., \& {Riess}, A. 2004, \apjl, 600, L103

\bibitem[{{Jannuzi} \& {Dey}(1999)}]{jannuzi_dey99}
{Jannuzi}, B.~T. \& {Dey}, A. 1999, in Astronomical Society of the Pacific
  Conference Series, Vol. 191, Photometric Redshifts and the Detection of High
  Redshift Galaxies, ed. {R.~Weymann, L.~Storrie-Lombardi, M.~Sawicki, \&
  R.~Brunner}, 111--+

\bibitem[{{Kajisawa} {et~al.}(2010){Kajisawa}, {Ichikawa}, {Yamada},
  {Uchimoto}, {Yoshikawa}, {Akiyama}, \& {Onodera}}]{kajisawa10}
{Kajisawa}, M., {Ichikawa}, T., {Yamada}, T., {Uchimoto}, Y.~K., {Yoshikawa},
  T., {Akiyama}, M., \& {Onodera}, M. 2010, ArXiv e-prints

\bibitem[{{Kennicutt}(1998)}]{kennicutt98}
{Kennicutt}, Jr., R.~C. 1998, \araa, 36, 189

\bibitem[{{Keres} {et~al.}(2008){Keres}, {Katz}, {Fardal}, {Dave}, \&
  {Weinberg}}]{keres08}
{Keres}, D., {Katz}, N., {Fardal}, M., {Dave}, R., \& {Weinberg}, D.~H. 2008,
  ArXiv e-prints

\bibitem[{{Kolatt} {et~al.}(1999){Kolatt}, {Bullock}, {Somerville}, {Sigad},
  {Jonsson}, {Kravtsov}, {Klypin}, {Primack}, {Faber}, \& {Dekel}}]{kolatt99}
{Kolatt}, T.~S., {Bullock}, J.~S., {Somerville}, R.~S., {Sigad}, Y., {Jonsson},
  P., {Kravtsov}, A.~V., {Klypin}, A.~A., {Primack}, J.~R., {Faber}, S.~M., \&
  {Dekel}, A. 1999, \apjl, 523, L109

\bibitem[{{Kron}(1980)}]{kron80}
{Kron}, R.~G. 1980, \apjs, 43, 305

\bibitem[{{Lee} {et~al.}(2009){Lee}, {Giavalisco}, {Conroy}, {Wechsler},
  {Ferguson}, {Somerville}, {Dickinson}, \& {Urry}}]{lee09}
{Lee}, K., {Giavalisco}, M., {Conroy}, C., {Wechsler}, R.~H., {Ferguson},
  H.~C., {Somerville}, R.~S., {Dickinson}, M.~E., \& {Urry}, C.~M. 2009, \apj,
  695, 368

\bibitem[{{Lee} {et~al.}(2006){Lee}, {Giavalisco}, {Gnedin}, {Somerville},
  {Ferguson}, {Dickinson}, \& {Ouchi}}]{lee06}
{Lee}, K.-S., {Giavalisco}, M., {Gnedin}, O.~Y., {Somerville}, R.~S.,
  {Ferguson}, H.~C., {Dickinson}, M., \& {Ouchi}, M. 2006, \apj, 642, 63

\bibitem[{{Leitherer} {et~al.}(1999){Leitherer}, {Schaerer}, {Goldader},
  {Gonz{\'a}lez Delgado}, {Robert}, {Kune}, {de Mello}, {Devost}, \&
  {Heckman}}]{leitherer99}
{Leitherer}, C., {Schaerer}, D., {Goldader}, J.~D., {Gonz{\'a}lez Delgado},
  R.~M., {Robert}, C., {Kune}, D.~F., {de Mello}, D.~F., {Devost}, D., \&
  {Heckman}, T.~M. 1999, \apjs, 123, 3

\bibitem[{{Madau}(1995)}]{madau95}
{Madau}, P. 1995, \apj, 441, 18

\bibitem[{{Madau} {et~al.}(1996){Madau}, {Ferguson}, {Dickinson}, {Giavalisco},
  {Steidel}, \& {Fruchter}}]{madau96}
{Madau}, P., {Ferguson}, H.~C., {Dickinson}, M.~E., {Giavalisco}, M.,
  {Steidel}, C.~C., \& {Fruchter}, A. 1996, \mnras, 283, 1388

\bibitem[{{Magdis} {et~al.}(2010){Magdis}, {Rigopoulou}, {Huang}, \&
  {Fazio}}]{magdis10}
{Magdis}, G.~E., {Rigopoulou}, D., {Huang}, J., \& {Fazio}, G.~G. 2010, \mnras,
  401, 1521

\bibitem[{{Mannucci} {et~al.}(2009){Mannucci}, {Cresci}, {Maiolino}, {Marconi},
  {Pastorini}, {Pozzetti}, {Gnerucci}, {Risaliti}, {Schneider}, {Lehnert}, \&
  {Salvati}}]{mannucci09}
{Mannucci}, F., {Cresci}, G., {Maiolino}, R., {Marconi}, A., {Pastorini}, G.,
  {Pozzetti}, L., {Gnerucci}, A., {Risaliti}, G., {Schneider}, R., {Lehnert},
  M., \& {Salvati}, M. 2009, \mnras, 398, 1915

\bibitem[{{Maraston} {et~al.}(2010){Maraston}, {Pforr}, {Renzini}, {Daddi},
  {Dickinson}, {Cimatti}, \& {Tonini}}]{maraston10}
{Maraston}, C., {Pforr}, J., {Renzini}, A., {Daddi}, E., {Dickinson}, M.,
  {Cimatti}, A., \& {Tonini}, C. 2010, ArXiv e-prints

\bibitem[{{Marchesini} {et~al.}(2009){Marchesini}, {van Dokkum}, {F{\"o}rster
  Schreiber}, {Franx}, {Labb{\'e}}, \& {Wuyts}}]{marchesini09}
{Marchesini}, D., {van Dokkum}, P.~G., {F{\"o}rster Schreiber}, N.~M., {Franx},
  M., {Labb{\'e}}, I., \& {Wuyts}, S. 2009, \apj, 701, 1765

\bibitem[{{Marchesini} {et~al.}(2010){Marchesini}, {Whitaker}, {Brammer}, {van
  Dokkum}, {Labbe}, {Muzzin}, {Quadri}, {Kriek}, {Lee}, {Rudnick}, {Franx},
  {Illingworth}, \& {Wake}}]{marchesini10}
{Marchesini}, D., {Whitaker}, K.~E., {Brammer}, G., {van Dokkum}, P.~G.,
  {Labbe}, I., {Muzzin}, A., {Quadri}, R.~F., {Kriek}, M., {Lee}, K.,
  {Rudnick}, G., {Franx}, M., {Illingworth}, G.~D., \& {Wake}, D. 2010, ArXiv
  e-prints

\bibitem[{{Martini} \& {Weinberg}(2001)}]{martini01}
{Martini}, P. \& {Weinberg}, D.~H. 2001, \apj, 547, 12

\bibitem[{{McLure} {et~al.}(2009){McLure}, {Cirasuolo}, {Dunlop}, {Foucaud}, \&
  {Almaini}}]{mclure09}
{McLure}, R.~J., {Cirasuolo}, M., {Dunlop}, J.~S., {Foucaud}, S., \& {Almaini},
  O. 2009, \mnras, 395, 2196

\bibitem[{{Meurer} {et~al.}(1999){Meurer}, {Heckman}, \& {Calzetti}}]{meurer99}
{Meurer}, G.~R., {Heckman}, T.~M., \& {Calzetti}, D. 1999, \apj, 521, 64

\bibitem[{{Mihos} \& {Hernquist}(1996)}]{mihos96}
{Mihos}, J.~C. \& {Hernquist}, L. 1996, \apj, 464, 641

\bibitem[{{Noeske} {et~al.}(2007){Noeske}, {Weiner}, {Faber}, {Papovich},
  {Koo}, {Somerville}, {Bundy}, {Conselice}, {Newman}, {Schiminovich}, {Le
  Floc'h}, {Coil}, {Rieke}, {Lotz}, {Primack}, {Barmby}, {Cooper}, {Davis},
  {Ellis}, {Fazio}, {Guhathakurta}, {Huang}, {Kassin}, {Martin}, {Phillips},
  {Rich}, {Small}, {Willmer}, \& {Wilson}}]{noeske07}
{Noeske}, K.~G., {Weiner}, B.~J., {Faber}, S.~M., {Papovich}, C., {Koo}, D.~C.,
  {Somerville}, R.~S., {Bundy}, K., {Conselice}, C.~J., {Newman}, J.~A.,
  {Schiminovich}, D., {Le Floc'h}, E., {Coil}, A.~L., {Rieke}, G.~H., {Lotz},
  J.~M., {Primack}, J.~R., {Barmby}, P., {Cooper}, M.~C., {Davis}, M., {Ellis},
  R.~S., {Fazio}, G.~G., {Guhathakurta}, P., {Huang}, J., {Kassin}, S.~A.,
  {Martin}, D.~C., {Phillips}, A.~C., {Rich}, R.~M., {Small}, T.~A., {Willmer},
  C.~N.~A., \& {Wilson}, G. 2007, \apjl, 660, L43

\bibitem[{{Oke} \& {Gunn}(1983)}]{oke83}
{Oke}, J.~B. \& {Gunn}, J.~E. 1983, \apj, 266, 713

\bibitem[{{Ouchi} {et~al.}(2005){Ouchi}, {Hamana}, {Shimasaku}, {Yamada},
  {Akiyama}, {Kashikawa}, {Yoshida}, {Aoki}, {Iye}, {Saito}, {Sasaki},
  {Simpson}, \& {Yoshida}}]{ouchi05}
{Ouchi}, M., {Hamana}, T., {Shimasaku}, K., {Yamada}, T., {Akiyama}, M.,
  {Kashikawa}, N., {Yoshida}, M., {Aoki}, K., {Iye}, M., {Saito}, T., {Sasaki},
  T., {Simpson}, C., \& {Yoshida}, M. 2005, \apjl, 635, L117

\bibitem[{{Pannella} {et~al.}(2009){Pannella}, {Carilli}, {Daddi}, {McCracken},
  {Owen}, {Renzini}, {Strazzullo}, {Civano}, {Koekemoer}, {Schinnerer},
  {Scoville}, {Smol{\v c}i{\'c}}, {Taniguchi}, {Aussel}, {Kneib}, {Ilbert},
  {Mellier}, {Salvato}, {Thompson}, \& {Willott}}]{pannella09}
{Pannella}, M., {Carilli}, C.~L., {Daddi}, E., {McCracken}, H.~J., {Owen},
  F.~N., {Renzini}, A., {Strazzullo}, V., {Civano}, F., {Koekemoer}, A.~M.,
  {Schinnerer}, E., {Scoville}, N., {Smol{\v c}i{\'c}}, V., {Taniguchi}, Y.,
  {Aussel}, H., {Kneib}, J.~P., {Ilbert}, O., {Mellier}, Y., {Salvato}, M.,
  {Thompson}, D., \& {Willott}, C.~J. 2009, \apjl, 698, L116

\bibitem[{{Papovich} {et~al.}(2001){Papovich}, {Dickinson}, \&
  {Ferguson}}]{papovich01}
{Papovich}, C., {Dickinson}, M., \& {Ferguson}, H.~C. 2001, \apj, 559, 620

\bibitem[{{Papovich} {et~al.}(2010){Papovich}, {Finkelstein}, {Ferguson},
  {Lotz}, \& {Giavalisco}}]{papovich10}
{Papovich}, C., {Finkelstein}, S.~L., {Ferguson}, H.~C., {Lotz}, J.~M., \&
  {Giavalisco}, M. 2010, ArXiv e-prints

\bibitem[{{Pentericci} {et~al.}(2007){Pentericci}, {Grazian}, {Fontana},
  {Salimbeni}, {Santini}, {de Santis}, {Gallozzi}, \&
  {Giallongo}}]{pentericci07}
{Pentericci}, L., {Grazian}, A., {Fontana}, A., {Salimbeni}, S., {Santini}, P.,
  {de Santis}, C., {Gallozzi}, S., \& {Giallongo}, E. 2007, \aap, 471, 433

\bibitem[{{Quadri} {et~al.}(2007){Quadri}, {van Dokkum}, {Gawiser}, {Franx},
  {Marchesini}, {Lira}, {Rudnick}, {Herrera}, {Maza}, {Kriek}, {Labb{\'e}}, \&
  {Francke}}]{quadri07}
{Quadri}, R., {van Dokkum}, P., {Gawiser}, E., {Franx}, M., {Marchesini}, D.,
  {Lira}, P., {Rudnick}, G., {Herrera}, D., {Maza}, J., {Kriek}, M.,
  {Labb{\'e}}, I., \& {Francke}, H. 2007, \apj, 654, 138

\bibitem[{{Reddy} {et~al.}(2010){Reddy}, {Erb}, {Pettini}, {Steidel}, \&
  {Shapley}}]{reddy10}
{Reddy}, N.~A., {Erb}, D.~K., {Pettini}, M., {Steidel}, C.~C., \& {Shapley},
  A.~E. 2010, \apj, 712, 1070

\bibitem[{{Reddy} \& {Steidel}(2009)}]{reddy09}
{Reddy}, N.~A. \& {Steidel}, C.~C. 2009, \apj, 692, 778

\bibitem[{{Reddy} {et~al.}(2006){Reddy}, {Steidel}, {Fadda}, {Yan}, {Pettini},
  {Shapley}, {Erb}, \& {Adelberger}}]{reddy06a}
{Reddy}, N.~A., {Steidel}, C.~C., {Fadda}, D., {Yan}, L., {Pettini}, M.,
  {Shapley}, A.~E., {Erb}, D.~K., \& {Adelberger}, K.~L. 2006, \apj, 644, 792

\bibitem[{{Renzini}(2009)}]{renzini09}
{Renzini}, A. 2009, \mnras, 398, L58

\bibitem[{{Sawicki} \& {Thompson}(2006)}]{sawicki06}
{Sawicki}, M. \& {Thompson}, D. 2006, \apj, 642, 653

\bibitem[{{Shapley} {et~al.}(2001){Shapley}, {Steidel}, {Adelberger},
  {Dickinson}, {Giavalisco}, \& {Pettini}}]{shapley01}
{Shapley}, A.~E., {Steidel}, C.~C., {Adelberger}, K.~L., {Dickinson}, M.,
  {Giavalisco}, M., \& {Pettini}, M. 2001, \apj, 562, 95

\bibitem[{{Shapley} {et~al.}(2005){Shapley}, {Steidel}, {Erb}, {Reddy},
  {Adelberger}, {Pettini}, {Barmby}, \& {Huang}}]{shapleyetal05}
{Shapley}, A.~E., {Steidel}, C.~C., {Erb}, D.~K., {Reddy}, N.~A., {Adelberger},
  K.~L., {Pettini}, M., {Barmby}, P., \& {Huang}, J. 2005, \apj, 626, 698

\bibitem[{{Shapley} {et~al.}(2003){Shapley}, {Steidel}, {Pettini}, \&
  {Adelberger}}]{shapley03}
{Shapley}, A.~E., {Steidel}, C.~C., {Pettini}, M., \& {Adelberger}, K.~L. 2003,
  \apj, 588, 65

\bibitem[{{Somerville} {et~al.}(2001){Somerville}, {Primack}, \&
  {Faber}}]{somerville01}
{Somerville}, R.~S., {Primack}, J.~R., \& {Faber}, S.~M. 2001, \mnras, 320, 504

\bibitem[{{Stark} {et~al.}(2009){Stark}, {Ellis}, {Bunker}, {Bundy}, {Targett},
  {Benson}, \& {Lacy}}]{stark09}
{Stark}, D.~P., {Ellis}, R.~S., {Bunker}, A., {Bundy}, K., {Targett}, T.,
  {Benson}, A., \& {Lacy}, M. 2009, \apj, 697, 1493

\bibitem[{{Steidel} {et~al.}(1999){Steidel}, {Adelberger}, {Giavalisco},
  {Dickinson}, \& {Pettini}}]{steidel99}
{Steidel}, C.~C., {Adelberger}, K.~L., {Giavalisco}, M., {Dickinson}, M., \&
  {Pettini}, M. 1999, \apj, 519, 1

\bibitem[{{Vanzella} {et~al.}(2009){Vanzella}, {Giavalisco}, {Dickinson},
  {Cristiani}, {Nonino}, {Kuntschner}, {Popesso}, {Rosati}, {Renzini}, {Stern},
  {Cesarsky}, {Ferguson}, \& {Fosbury}}]{vanzella09}
{Vanzella}, E., {Giavalisco}, M., {Dickinson}, M., {Cristiani}, S., {Nonino},
  M., {Kuntschner}, H., {Popesso}, P., {Rosati}, P., {Renzini}, A., {Stern},
  D., {Cesarsky}, C., {Ferguson}, H.~C., \& {Fosbury}, R.~A.~E. 2009, \apj,
  695, 1163

\end{thebibliography}

\end{document}